\documentclass[12pt, onecolumn,draftcls]{IEEEtran}
\usepackage{enumerate}
\usepackage{amsmath,amssymb,bm,amsfonts}
\usepackage{graphics,graphicx,epsfig,subfigure}
\usepackage{algorithm,algorithmic}
\usepackage{theorem}
\usepackage{threeparttable}
\usepackage{stfloats}
\usepackage{cite}
\usepackage{multirow}
\usepackage{mathrsfs}
\usepackage{amsmath,bm,bbm}
\usepackage{color}
\usepackage{multirow}
\usepackage{subeqnarray}
\usepackage{cases}
\usepackage{colortbl}
\usepackage{wasysym}
\usepackage[dvipsnames]{xcolor}
\usepackage{url}

\newtheorem{mytheorem}{Theorem}

\newtheorem{myproof}{Proof}
\newtheorem{mydef}{Definition}
\newtheorem{myproperty}{Property}
\begin{document}
\newcommand*{\QEDA}{\hfill\ensuremath{\blacksquare}}
\def\arrow{{\rightarrow}}
\def\N{{\mathcal{N}}}
\def\B{{\mathcal{B}}}
\def\E{{\mathcal{E}}}
\def\I{{\mathcal{I}}}
\def\diag{{\textrm{diag}}}
\def\i {{ -i}}
\def\ci{\perp\!\!\!\perp} 
\newcommand\independent{\protect\mathpalette{\protect\independenT}{\perp}} 
\def\independenT#1#2{\mathrel{\rlap{$#1#2$}\mkern2mu{#1#2}}}

\title{Proactive Doppler Shift Compensation in Vehicular Cyber-Physical Systems}
\author{Jian Du, Xue Liu and  Lei Rao
\thanks{Jian Du and Xue Liu are with the School  of Computer Science, McGill University (e-mail: dujianeee@gmail.com, xueliu@cs.mcgill.ca).}
\thanks{ Lei Rao is with General Motors, United States  (e-mail: lei.rao@gm.com).}
}
\maketitle

\IEEEpeerreviewmaketitle
\begin{abstract}
In vehicular cyber-physical systems (CPS), safety information, including vehicular speed and location information, is shared among vehicles via wireless waves at specific frequency. This helps control vehicle to alleviate traffic congestion and road accidents.
However, Doppler shift existing between vehicles with high relative speed
causes an apparent frequency shift for the received wireless wave, which
consequently decreases the reliability of the recovered safety information and jeopardizes the safety of vehicular CPS.
Passive confrontation of Doppler shift at the receiver side is not applicable due to multiple Doppler shifts at each receiver.
In this paper, we provide a proactive Doppler shift compensation algorithm based on the probabilistic  graphical model.
Each vehicle pre-compensates its carrier frequency individually so that there is no frequency shift from the desired carrier frequency between each pair of transceiver.
The pre-compensated offset for each vehicle is computed in a distributed fashion in order to be adaptive to the distributed and dynamic topology of vehicular CPS.
Besides, the updating procedure is designed in a broadcasting fashion to reduce communication burden.
It is rigorously proved that the proposed algorithm is convergence guaranteed even for systems with packet drops and random communication delays.
Simulations based on real map and transportation data verify the accuracy and convergence property of the proposed algorithm.
It is shown that this method achieves almost the optimal frequency compensation accuracy with an error approaching the Cram\'{e}r-Rao lower bound.
\end{abstract}

%
%
\section{Introduction}
\subsection{Context and Motivation}
Developing vehicles from a
purely physical system based on the laws of mechanics and
chemistry, to a more sophisticated and intelligent cyber physical system (CPS) with functions of communication and control
is a promising direction to enhance traffic safety and efficiency.
In vehicular CPS,
vehicle safety information, e.g., speed, location, and acceleration, are shared with high reliability among different vehicles,
so that cooperative vehicle control \cite{Murray-control} can be applied to improve the driving safety and alleviate the traffic congestion.
The U.S. Department of Transportation estimates that vehicular CPS could help address up to $81$ percent of crash scenarios with unimpaired drivers, preventing tens of thousands of automobile crashes every year \cite{CPS-Connect}.

Doppler shift, which is the perceived change in frequency of wave emitted by a source which is moving
relative to an observer, exists among vehicles
due to their mobility.
Since safety information is shared via wireless waves at specific frequency,
the received waves would be moved from the desired frequency due to
Doppler shift, which consequently decreases the reliability of the recovered safety information and thus jeopardizes the safety of vehicular CPS.
\begin{figure}[b]
  \centering
\epsfig{file=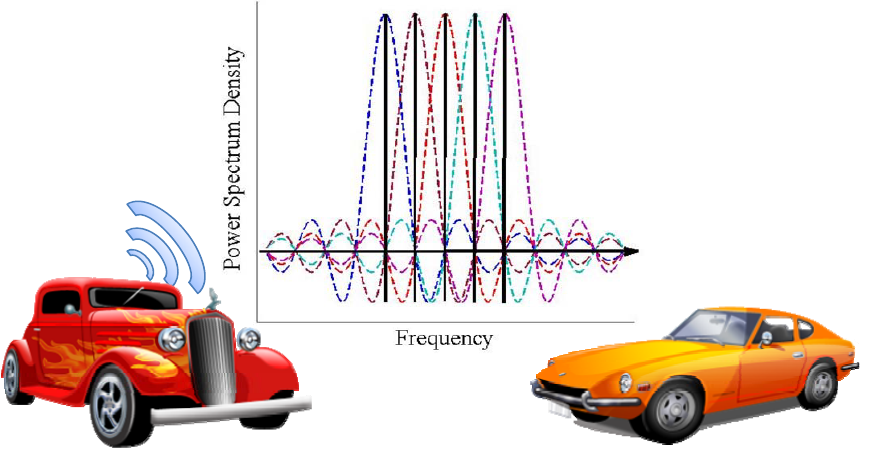, width=3.2in}
\caption{Orthogonal sub-carriers are utilized for vehicle safety information sharing.
At the frequency that one sub-carrier takes its peak value, all other sub-carriers are zero.
Hence, sampling at frequencies that take peak values is important for  safety information recovering at the receiver.}
\label{veh-OFDM}
\end{figure}

More specifically,
safety information is shared via dedicated short range communications (DSRC) \cite{Kenney:2011Procee} and IEEE 802.11p protocal, which
utilize orthogonal frequency division multiplexing (OFDM) carrier waves to improve spectrum efficiency on 5.9GHz band.
{Although IEEE 802.11p is considered  the de facto standard for on-the-road communications \cite{2, wang2016smart}, researchers, manufacturers and stakeholder indeed have  started to investigate the usability of Long Term Evolution (LTE), in which orthogonal frequency-division multiple access (OFDMA) is  adopted for multiplexing, to support vehicular communications.
Interesting readers please refer  \cite{1,2,3,4,5,6} and the references therein.
Another interesting point is
that several auto manufacturers are considering
solutions for communication in inter-vehicle communication environments.
As reported in \cite{2}, several original equipment manufacturers
 have  announced agreements
with cellular carriers to use equipment
from those specific carriers in their vehicles for
Internet access and other services. This entails
the use of a LTE modem
installed in cars and the use of LTE (or LTE-advanced)
networks of carriers for several services.
Moreover, recently, the Qualcomm Snapdragon automotive
development platform, which supports not only IEEE 802.11p  but also LTE  for dedicated short range communications (DSRC) \cite{Kenney:2011Procee}, was
released \cite{5} to enable auto manufactures, suppliers and developers
to rapidly innovate, test and deploy vehicular applications.}
The OFDMA signal can be described as a set of closely spaced  frequency division sub-carriers.
In the frequency domain, each  sub-carrier is in sinc function form and sub-carriers are allocated to different users.
To deliver the safety information, each sub-carrier
is modulated with a conventional digital modulation scheme (such as QPSK, 16-QAM, etc.) and will be recovered at the receiver.
As shown in Fig. \ref{veh-OFDM},
though the side lobes of different sinc signals overlap with each other,
at the peak of each sinc signal, all other sinc signals are zero.
This fact guarantees that there is no inter-carrier interference if the receiver samples exactly at these peak  locations.
The peak locations may deviate from the pre-defined frequency due to Doppler shift.
Because the relative speed between vehicles may be high and results in large Doppler shift, the sampled frequency can not be exactly at the individual peak.
Therefore, the sampled value contains not only the desired sub-carrier information but also those from other sub-carriers as interferences.
Doppler shift would destroy the orthogonal property of different sub-carriers, and it is shown by theoretical analysis and verified by experiments that Doppler shift leads to
degradation of system capacity and bit error rate \cite{V2V-PIEEE}.

Doppler shift compensation has been studied for one pair transceiver with centralized processing method. However,
existing solutions \cite{JWChen, YiqingZhou,zhou} cannot be applied to  vehicular CPS due to the following difficulties:
1) For communications between one pair transceiver,
the frequency shift can be estimated and compensated at the receiver side \cite{JWChen,YiqingZhou,zhou}.
{In vehicular CPS, however, at each receiver
safety information from different vehicles arrives at the same time on different sub-carriers,
therefore it is necessary
to adjust the sampling frequency to compensate frequency offsets caused by different Doppler frequency shifts.
Let $i$ and $j$ denote the transmitter and receiver respectively.
Via training sequence based method \cite{cai2010cfo}, each receiver  first obtains  Doppler shift estimates $f_{i,j}$ and then adjusts the sampling frequency by multiplying $\exp(-\iota \frac{2\pi f_{i,j}t}{M})$ on the $t^{\textrm{th}}$ sample of the received baseband signal \cite{JWChen}, with $\iota $ denoting the imaginary unit and $M$
denoting the total parallel subcarriers adopted in the OFDMA scheme.
It is evident that when receiver $j$ receives data from more than one transmitters, it cannot compensate all the Doppler shifts since the received signal is a linear superposition of signals from different transmitters.
Therefore it is impossible to adjust the sampling frequency for compensating different frequency shifts.}
2) Due to the moving property of vehicles, the network topology is highly dynamic, and vehicles may also randomly join and leave the network.
Therefore, a distributed algorithm for frequency shift compensation is more suitable than the centralized method to  adapt the varying network topology.
3) As vehicular CPS may have high density and transmit large volume of data \cite{chen2015signal, chen2015discrete}, it is prone to resulting in broadcasting storm
\cite{BroadcastinginVanet, LowCongestionControl},
and thus, an algorithm with communication overhead linearly scaling with the vehicle density is desired.
{To solve the above challenges, distributed  algorithm is proposed and is adopted after each receiver obtains the Doppler shifts estimate with the  training based method.}
\subsection{Contributions and Organization of the paper}
To address above challenges, we propose proactive Doppler shift compensation algorithm based on the probabilistic graphical model.
{We assume data are transmitted frame by frame.
Each time, when the transmitter sends a data frame out, it is reasonable to assume that Doppler shift for this data frame  is a fixed constant due to the fact that the time duration for each data frame is much smaller than the vehicle speed change.
The proposed algorithm compensates Doppler shift for each data fame in a distributed fashion.}
We study this algoirthm from both algorithm design and theoretical analysis perspectives.

From the algorithm design perspective, we construct a probabilistic graphical model
to reveal the conditional independence structure of Doppler shifts in vehicular CPS.
Though
the classical belief propagation (BP) algorithm  \cite{du2016convergence} can be applied to distributed frequency shift compensation,
the number of messages involved in BP algorithm at each iteration  grows quadratically as the number of vehicles increases, leading to information network congestion.
To overcome this problem, we propose a novel distributed algorithm named as linear scaling belief propagation (LSBP) for its linear scalability to network density.
We apply LSBP to a vehicular network with arbitrary topologies and with potential packet drops as well as random transmission and processing delays. It is shown that the total number of messages at each iteration simply equals to the number of vehicles.

From the theoretical analysis perspective, the convergence properties are analyzed for LSBP.
Note that though BP has gained great success in many applications,  it is found that BP may diverge if the network topology contains circles,
and the necessary and sufficient convergence condition is still an open problem.
Thus, BP is not reliable for vehicular CPS.
In contrast, the analytical analysis of the proposed LSBP algorithm shows that LSBP is convergence guaranteed for arbitrary vehicular network topology and is robust to
packet drops and random delays.
Besides, even with different initial values, the LSBP converges to a unique point.
The above theoretical analysis is also verified by simulations, and it is shown that the proposed LSBP algorithm converges quickly with the estimation mean-square-error (MSE) approaching the Cram\'{e}r-Rao lower bound (CRLB) even under dynamic topologies.{
Previous works on distributed estimation \cite{du2013network,du2013distributed}  focus on static network and convergence of standard BP for distributed estimation is analyzed.
However, vehicular network is dynamic and may be very dense in certain area.
This paper proposes LSBP algorithm, in which
the updating procedure is designed in a broadcasting fashion
to reduce communication burden and convergence guaranteed property is analytically shown.}

The rest of the paper is organized as follows.
A motivating example is shown in Section~\ref{Example} to provide some intuitive insights.
The general model and problem formulation are introduced in Section~\ref{model}.
The distributed estimation algorism based on probabilistic graphical model is presented in Section~\ref{algorithm}.
In Section~\ref{analysis}, convergence property of the proposed algorithm is analytically proved.
Simulation results of proactive frequency shift compensation are illustrated in Section~\ref{simu}.
Concluding remarks are given in Section~\ref{conclusion}.

\textit{Notations}: Boldface uppercase and lowercase letters represent matrices and vectors, respectively.
$\mathbb{E}$ denotes the statistical expectation operator.
$\bm A^{-1}$ and $\bm A^T$ denote the inverse and the transpose of matrix $\bm A$, respectively.
Notation $\N(x; \mu,  P)$ stands for the probability density function (PDF) of a Gaussian random variable $x$ with mean $\mu$ and variance $P$.
Symbol $ \propto$ represents the linear scalar relationship between two real valued functions, and
$\diag\{\bm A\}$ refers to taking the diagonal element of $\bm A$.

%
\section{Motivating Example}\label{Example}
Combating Doppler shift is also a problem in nature.
Certain species of bats, who can produce constant frequency echolocation
calls, compensate for the Doppler shift by lowering their call frequency as they approach a target.
This keeps the returning echo in the same frequency range of the normal echolocation call.
This dynamic frequency modulation  was discovered by Hans Schnitzler
in 1989 \cite{Nature-Bat}.
Inspired by this example, we propose proactive frequency shift compensation at each transmitter to mitigate each pair of Doppler shift in vehicular networks.
In the following we give a three-vehicle example to explain the main idea.

A vehicular CPS consisting of three vehicles as shown in Fig. 2(a) is considered in this motivating example.
It is assumed that vehicles 1, 2 and 3 are within the communication range of each other.
Vehicle 3 receives safety information broadcasted by vehicles 1 and 2 simultaneously.
Doppler shift between vehicles 1 and 3
is designated by $f_{1,3}$, and Doppler shift between vehicles 2 and 3 is $f_{2,3}$.
Due to different relative velocities between vehicles, we have $f_{1,3}\neq f_{2,3}$.
Therefore, if vehicle $3$ compensates the frequency shift by $f_{1,3}$, there is still a frequency shift mismatch between vehicles 2 and 3.
However, if each vehicle can proactively compensate by certain frequency amount before sending safety information, it is possible to mitigate the frequency offset.
For instance, as shown in Fig. 2(b),
let the pre-compensated frequency shift at vehicles 1, 2 and 3 be $f_1$, $f_2$, and $f_3$, respectively.
Then if $f_1$, $f_2$, and $f_3$ satisfy
$f_{1} + f_{2} = f_{1,2}$, $f_{1} + f_{3} = f_{1,3}$, and
$f_{2} + f_{3} = f_{2, 3}$,
there will be no frequency offset for each received signal at any vehicle.

How to obtain $f_i$ ($f_1$, $f_2$, and $f_3$ in this example) is not an easy problem due to the following challenges:
\begin{itemize}
\item
$f_{i,j}$ ($f_{1,2}$, $f_{1,3}$ and $f_{2,3}$ in this example) cannot be exactly known since
the true relative frequency shift $f_{i,j}$ can only be approximated via statistical estimate \cite{JWChen} or measurements.
\item
Since the number of vehicles is large, the centralized method, which requires the information of all $f_{i,j}$ and the network topology of vehicular CPS, is difficult to be implemented.
Hence, distributed estimation which only involves local computation at each vehicle is desired.
\item
The distributed method needs additional information exchange between neighbors for iterative updating, and the number of messages needed for updating should be linear scaling with the vehicle density.
\item
The distributed estimation algorithm should also be adaptive to the dynamic topology of vehicular CPS.
Besides, convergence of the algorithm has to be guaranteed.
\end{itemize}

\begin{figure}[t]
  \centering
\mbox{
\subfigure[]
{\epsfig{file=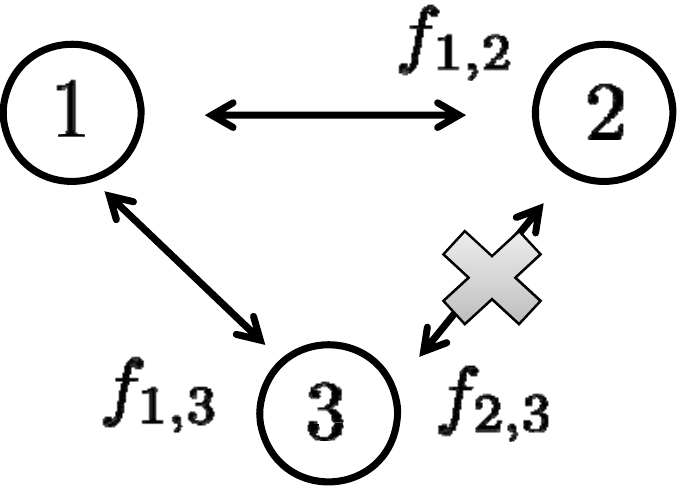,width=1.4in}}    }\label{eg1}
\mbox{
\subfigure[]
{\epsfig{figure=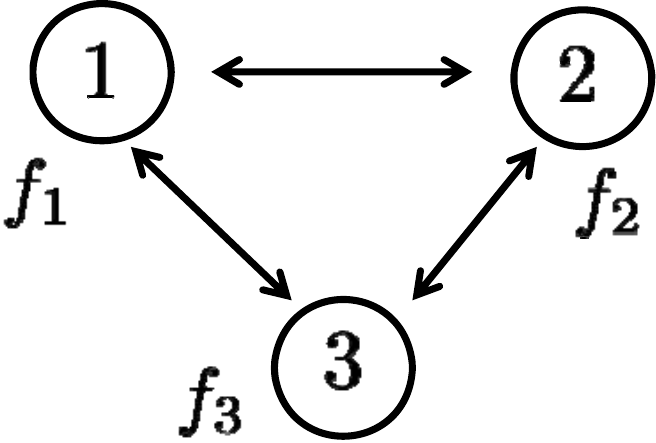,width=1.4in}}\label{eg2}       }
\caption{Comparison of passive and proactive Doppler shift compensation methods.
(a) Passive method: At vehicle $3$,  $f_{1,2}$  and  $f_{2,3}$ cannot be compensated at the same time.
(b) Proactive method: Pre-compensating $f_i$ at each vehicle  results in no relative Doppler shift between each pair of communication link.}
  \label{eg}
\end{figure}


\section{Problem Formulation and Modeling
}\label{model}

\subsection{Optimal Performance}
The interaction topology of a vehicular CPS is represented
by an undirected graph
$\mathcal{G}=(\mathcal{V},\mathcal{E})$,
where $\mathcal{V}=\{1,\ldots, N\}$ is the set of vehicles, and $\mathcal{E}\subseteq \mathcal{V}\times \mathcal{V}$ is the set of communication links.
Although we assume the vertices $\mathcal{V}$ to be fixed and indexed in a certain order, the mathematical theory that follows does not change if the names of the vertices are rearranged.
Vehicles within communication range of each other are regarded as neighbors, and
neighbors of vehicle $i$
are denoted by $\mathcal{B}(i)\triangleq \{j\in \mathcal{V}| (i,j)\in \mathcal{E}\}$.
To model the communication link failures, $\mathcal{G}$ is assumed to be a Bernoulli network:  at each communication, a network link is
active with some probability;  network links may have different link probabilities; and links fail or are
alive independently of each other.

Let $f_{i,j}$ be the Doppler shift between $i$ and $j$, then the pre-compensated frequency shift at vehicle $i$ and at vehicle $j$, i.e., $f_{i}$ and $f_{j}$, should satisfy
$f_i + f_j = f_{i,j}$.
In practice, we can only obtain the measurement or estimate \cite{DopplorSpeed,JWChen} of $f_{i,j}$, denoted as $r_{i,j}$, between neighboring vehicles $\{i,j\}\in \E$.
Thus, we have
\begin{equation} \label{linearSPEED}
r_{i,j}
=  f_i +  f_j + n_{i,j},
\end{equation}
where $n_{i,j}$ is the estimation error.
{It is known that  the maximum
likelihood  estimates of $n_{i,j}$ is asymptotically Gaussian distributed \cite{JWChen}, that is,
$n_{i,j}\sim \mathcal{N}( n_{i,j};  0, \sigma^2_{i,j})$.}
Let $\hat{f}_i$ denote the estimate of $f_i$.
The estimation MSE, defined as $\mathbb{E}\{(\hat{f}_i-f_i)^2\}$, is used to evaluate the performance of the estimator with the lower bound of MSE given as the performance benchmark.
Define $\bm f=[f_2, f_3,\ldots, f_N]^T$
\footnote{$f_1$ is set as the reference frequency which can be arbitrary constant}
 and stack (\ref{linearSPEED}) with respect to all $i$ and $j$ into a matrix form, we obtain
\begin{equation}\label{blocklinear}
\bm r=
\left[\begin{array}{cc}
\bm a& \!\! \bm A
\end{array} \right]\!\!
\left[\begin{array}{c}
f_1\\ \bm f
\end{array} \right] + \bm n,
\end{equation}
where $\bm r$ is a vector containing $ r_{i,j}$ with ascending indices first on $i$ and then on $j$,
and $\bm n$ containing $ n_{i,j}$ with the indices $i$, $j$ ordered in the same way as in $\bm r$.
Then, $\bm n \sim \N(\bm n;\bm0,\bm R)$,
where $\bm R$ is a diagonal matrix with $\sigma^2_{i,j}$ as  diagonal elements which have the same order as $r_{i,j}$ in $\bm r$.
$\left[\begin{array}{cc}
\bm a_1& \!\! \bm A
\end{array} \right]$ is a matrix containing $0$ and $1$ to make (\ref{blocklinear}) hold for each $(i,j)\in \mathcal{E}$, and $\bm a_1$ is its first column.
Note that (\ref{blocklinear}) is a standard linear model, so the Cram\'{e}r-Rao lower bound (CRLB) of $\bm f$, which provides the lower bound of the achievable MSE of any unbiased estimator, can be easily computed as \cite{PointEst}
\begin{equation}\textrm{CRLB}(\bm {f})
=\diag\{\big(\bm A^T\bm R^{-1}\bm A \big)^{-1}\}.
\label{crb}
\end{equation}
The maximum likelihood  estimator
is the best linear unbiased estimator approaching CRLB for the linear model of (\ref{blocklinear}), and is  given by \cite{PointEst}
\begin{eqnarray}\label{Central}
\begin{split}
[\hat{f}_2,\ldots, \hat{f}_N]^T
&\triangleq
\arg \mathop {\max }\limits_{f_1,\ldots, f_N}
\N(\bm r-f_1\bm a; \bm A\bm{f}, \bm R) \\
&=
 ( \bm A^T\bm R^{-1}\bm A)^{-1}\bm A^T\bm R^{-1}(\bm r-f_1\bm a) .
\end{split}
\end{eqnarray}

Implementing (\ref{Central}), however, not only requires bringing all  $r_{i,j}$ and $\sigma_{i,j}^2$ to a central computing unit, but also needs the  topology of $\mathcal{G}$ to construct $\bm r$ and $\bm A$.
Thus, the maximum likelihood estimator is not scalable with network size, which causes heavy communication  burden by transmitting data from network border to control unit.
Besides,  (\ref{Central}) needs to be re-computed frequently due to the dynamic property of vehicular networks.
Therefore, distributed estimation, where each vehicle performs estimation with local information, sounds promising \cite{Yang}.
However, achieving the optimal MSE as in (\ref{crb}) in a distributed fashion without global information is challenging.
Leveraging statistical property of $\{f_i\}_{i\in \mathcal{V}}$ for
distributed algorithm design is one  promising direction.
We next introduce the probabilistic graphical model to reveal  conditional independence structure of Doppler shifts in vehicular CPS.

\subsection{Primer on Probabilistic Graphical Model}
\begin{figure}[t]
  \centering
{\epsfig{file=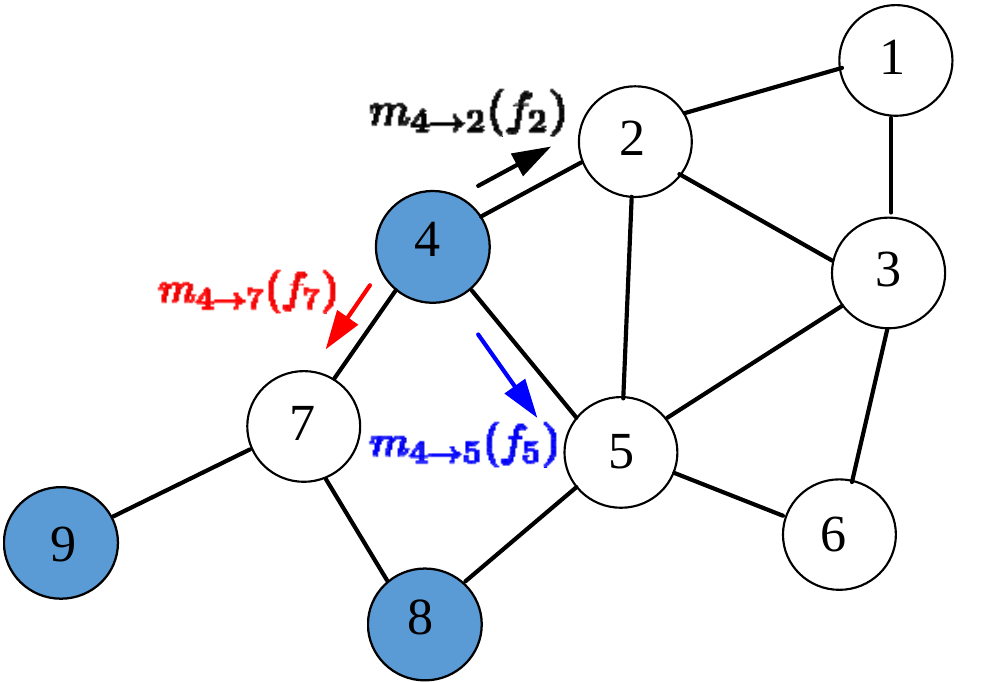,width=2.2in}}
\caption{An example of a graphical model for a vehicular CPS with $9$ vehicles.
Vertex $i$ in the graph corresponds to $f_i$ that needs to be pre-compensated on vehicle $i$.
The Markov blanket of node $7$
consists of the set of its neighbouring nodes $\{4,8,9\}$.
Besides,    messages from $4$ to its different neighbors are different in BP algorithm, and these messages  are denoted by different colors.}
\label{Graph}
\end{figure}

In a probabilistic graphical model, each vertex (node) represents
a random variable, and there are a set of edges joining
some pairs of vertices.
The graph gives a visual way of understanding the
joint distribution of an entire set of random variables on graph \cite{chen2015signal, pmudu}.
Fig. \ref{Graph} shows an example of a graphical model for a vehicular CPS with $9$ vehicles.
Vertex $i$ in the graph corresponds to $f_i$ that needs to pre-compensate on vehicle $i$.
According to (\ref{linearSPEED}), the  probabilistic relationship between $f_i$ and $f_j$ is captured by  $\N(r_{i,j};f_i+f_j,\sigma_{i,j}^2)$ and denoted on the graph by an edge linking these two variables.
Hence, the probabilistic graphical model has the same network topology as the vehicular CPS.
In this model, the absence of an edge between two vertices has
a special meaning: the corresponding random variables are conditionally
independent given one node's neighboring nodes.
These neighbors are known as \emph{Markov blanket}, i.e., $f_i$ and $f_j$ are conditional independent given all
$\{f_k\}_{k\in \mathcal{B}(i)}$ or all $\{f_k\}_{k\in \mathcal{B}(j)}$.
For example, as shown in Fig. \ref{Graph},
$f_7$ and $ f_2$ are conditional independent given $ \{f_4,f_8,f_9\}$.
Thus, it is possible to obtain the estimate of $f_7$  with the help of its neighbors, i.e., $\{4,8,9\}$, via message exchange.

\section{Distributed Algorithm Design}\label{algorithm}
In virtue of the conditional independence relationship between variables as revealed by the probabilistic  graphical model, distributed inference can be designed with only local information between neighbors.
In this section, leveraging the probabilistic graphical model, belief propagation (BP) algorithm is studied first for estimation of pre-compensated frequency shift. Inspired by BP, a distributed estimation algorithm named as linear scaling BP (LSBP), which has low communication overhead and is convergence guaranteed, is then proposed.
{Notice that communication scheme \cite{robust} that is robust to Doppler shift can be adopted for message exchange before Doppler shifts are compensated.}
\subsection{Belief Propagation Algorithm
}
With Gaussian belief propagation (BP) \cite{du2016convergence} algorithm for linear Gaussian model, at every iteration, each node sends a (different) message
to each of its neighbors and receives a message from each
neighbor.
The message from vehicle $j$ to vehicle $i$ is defined as
the product of the local function $\N(f_{i,j};f_i+f_j,\sigma_{i,j}^2)$ with messages received from all neighbors except $ i$,
and then maximized over all involved variables except $ f_i$.
Mathematically, it is defined as
\begin{equation}  \label{BPf2vs1}
\begin{split}
m^{(l)}_{j \arrow i}(f_i)
= &\mathop{\max}
\limits_{f_{j}}
\quad \N(f_{i,j};f_i+f_j,\sigma_{i,j}^2)\\
& \times \prod_{k\in\B(j)\setminus i} m^{(l-1)}_{k \arrow j}(f_j).
\end{split}
\end{equation}

The message $m^{(l)}_{j \arrow i}(f_i)$ is computed and exchanged among neighbors.
One possible  scheduling for message exchange is  that all  vehicles perform local computation and message exchange in parallel \cite{yang2017coding}.
In any round of message exchange, a belief of $ f_i$ can be computed at each vehicle $i$ locally,  as the product of all the incoming messages from neighbors, which is given by
\begin{equation} \label{BPbelief}
b^{(l)}(f_i)
 =  \prod_{ j\in\B(i)} m^{(l)}_{ j \arrow i}( f_i).
\end{equation}
The belief $b^{(l)}(f_i)$ serves as the approximation of $ \max_{\{f_1,\ldots, f_N\}\setminus f_i}
\N(\bm r; \bm A\bm{f}, \bm R)$.
Therefore, the estimate of $ f_i$ in the $l^{th}$ iteration can be computed by
\begin{equation}\label{est1}
\hat {f}_i^{(l)}
 =  \mathop{\max} \limits_{f_i} b^{(l)}(f_i).
\end{equation}

Note that the message and belief updating rules denoted by  (\ref{BPf2vs1})
and (\ref{BPbelief}) are naturally distributed:
message $m^{(l)}_{j \arrow i}(f_i)$ in (\ref{BPf2vs1})  is computed internally by $j$,
and then sent to  its neighbor $i$.
After receiving all the messages from its neighbors, $i$ computes the belief $b^{(l)}(f_i)$ according to (\ref{BPbelief}).

The message exchange between neighboring vehicles can be realized via local communications; however, as packet drops and random delays are the bottleneck of communication in vehicular CPS \cite{Bai:Mobi10}, their impact on message exchange should be addressed.
To do so, \emph{totally asynchronous scheduling} is adopted.
More specifically, each vehicle still performs message and belief computations at the individual predefined time, even when it doesn't receive newly updated messages from some of its neighbors.
This totally asynchronous scheduling is defined as follows.
\begin{mydef}
(Totally Asynchronous Scheduling):
The message available to  $j$ at time $l$ is
$m_{k\rightarrow j}^{(\tau_{k\rightarrow j}(l))}$ with $k\in \B(j)$, where $\tau_{k\rightarrow j}(l)$  satisfies $0\leq \tau_{k\rightarrow j}(l)\leq l$, and $\lim_{l\rightarrow +\infty}
\tau_{k\rightarrow j}(l) = +\infty $ for all $\{k,j\}\in \E$.
\end{mydef}

The physical meaning of the above definition is that, even though
packet drops and random delays may cause some updated messages failed to be received, local computation at each vehicle can still continue with part of the updated messages and part of the outdated messages received at the last iteration.
The outdated messages can  eventually be replaced by successfully received messages in the future updating.
Each vehicle $j$ keeps
a buffer with the most recently received messages from all its neighbors, i.e.,
$m^{(\tau_{k\rightarrow j}(l-1))}_{k \arrow j}(f_j)$ at iteration time $l$.
Therefore, under packet drops and random delays,
the outgoing message $m^{(l)}_{j \arrow i}(f_i)$ in (\ref{BPf2vs1}) can be computed as
\begin{equation}  \label{BPf2vs1-asy}
\begin{split}
m^{(l)}_{j \arrow i}(f_i)
=  \mathop{\max}
\limits_{f_{j}}
 \quad &\N(f_{i,j};f_i+f_j,\sigma_{i,j}^2)\\
& \times
\prod_{k\in\B(j)\setminus i} m^{(\tau_{k\rightarrow j}(l-1))}_{k \arrow j}(f_j).
\end{split}
\end{equation}
Similarly, the belief in (\ref{BPbelief}) can be computed as
\begin{equation} \label{BPbelief-asy}
b^{(l)}(f_i)
 =  \prod_{ j\in\B(i)} m^{(\tau_{k\rightarrow j}(l))}_{ j \arrow i}( f_i).
\end{equation}

\subsection{Message Computation for BP}
From (\ref{BPf2vs1-asy}) and (\ref{BPbelief-asy}), $m^{(l)}_{j \arrow  i}( f_i)$ and $b^{(l)}(f_i)$ are functions of variable $f_i$, and they represent the estimate of $f_i$ by $j$ and $i$, respectively.
As these messages are updated at each iteration,
explicit expressions of these messages are needed.
First, to facilitate the subsequent updating, the initial message is set to be in Gaussian function form  i.e., $\mathcal{N}( f_i;  \eta^{(0)}_{j\arrow i},  C_{j\arrow i}^{(0)})$.

Next, $m^{(l)}_{j \arrow  i}(  f_i)$ is computed.
Since $\N(f_{i,j};f_i+f_j,\sigma_{i,j}^2)$ is  a Gaussian function,
according to (\ref{BPf2vs1}),  $m^{(1)}_{j \arrow  i}(  f_i)$ is also a Gaussian
function, and by induction, it can be easily proved that
 $m^{(l)}_{j \arrow  i}(  f_i)$ in (\ref{BPf2vs1-asy})
 keeps Gaussian form for arbitrary $l$.
Therefore,
only its mean and variance   need to be transmitted for exchanging the message
$m^{(l)}_{j \arrow i}(f_i)$.

At this point, we can compute the messages at any iteration.
In general, for the $l^{\textrm{th}}$ ($l=2,3,\cdots$) round of message exchange, let the  available messages variance and mean at from $k$ to $j$ are $\big[C_{k\arrow j}^{(\tau_{k\rightarrow j}(l-1))}\big]^{-1}$ and $\eta^{(\tau_{k\rightarrow j}(l-1))}_{k\arrow j}$, respectivly.
Then, $j$ computes and transmits the outgoing messages to each of its neighbors individually.
After some
straightforward but tedious derivations, reciprocal of the message variance is given by
\begin{equation} \label{f2fC}
\big[C_{j\arrow i}^{(l)}\big]^{-1}
 =
\big[  \sigma_{i,j}^2 +
\big[
\!\sum_{k\in\mathcal{B}(j)\setminus i} \big[C_{k\arrow j}^{(\tau_{k\rightarrow j}(l-1))}\big]^{-1}\big]^{-1}
 \big]^{-1},
\end{equation}
and the message mean  is expressed as
\begin{eqnarray}\label{f2f2}
\eta^{(l)}_{j\arrow i}
&\!\!\!=\!\!\!&
\Big\{r_{i,j}+\big[\sum_{k\in\mathcal{B}(j)\setminus i} \big[ C_{k\arrow j}^{(\tau_{k\rightarrow j}(l-1))}\big]^{-1}\big]^{-1}\\\nonumber
&& \times
\big[
\sum_{k\in\mathcal{B}(j)\setminus i}
\big[ C_{k\arrow j}^{(\tau_{k\rightarrow j}(l-1))}\big]^{-1}\eta^{(\tau_{k\rightarrow j}(l-1))}_{k\arrow j}\big]\Big\}.
\end{eqnarray}

Due to packet drops and random delays, $\big[C_{j\arrow i}^{(l)}\big]^{-1}$
and $\eta^{(l)}_{j\arrow i}$, may or may not be successfully received by $i$ for updating
$b^{(l)}(f_i)$.
Following Definition 1, $ \big[  C_{j\arrow i}^{(\tau_{j\rightarrow i}(l))}\big]^{-1}$
and $ \eta^{(\tau_{k\rightarrow j}(l-1))}_{j\arrow i}$ are used to denote the available information of $i$ at iteration $l$, then $i$ can compute the BP estimates
via (\ref{BPbelief-asy}), which can be easily shown to be
$b_{i}^{(l)}( f_i)\propto  \mathcal{N}( f_i|\mu_i^{(l)},  P_{i}^{(l)})$,
with
\begin{equation} \label{beliefP}
 \big[  P_{i}^{(l)}\big]^{-1} = \sum_{j\in\B(i)}\big[  C_{j\arrow i}^{(\tau_{j\rightarrow i}(l))}\big]^{-1},
\end{equation}
and
\begin{equation}\label{beliefu}
\begin{split}
\hat {f}_i^{(l)}
&\triangleq
\mu_i^{(l)}
= P_{i}^{(l)} \sum_{j\in\B(i)}
  \big[  C_{j\arrow i}^{(\tau_{k\rightarrow j}(l-1))}\big]^{-1}  \eta^{(\tau_{k\rightarrow j}(l-1))}_{j\arrow i}.
 \end{split}
\end{equation}

Let $l_{\textrm{max}}$ denote the maximum updating times for each vehicle, and the algorithm terminates
when the maximum number of iteration $l_{\textrm{max}}$ is reached, or when $\Delta_i\triangleq\|\hat{v}_i^{(l)}-\hat{v}_i^{(l-1)}\|<th$, where $th$ is a threshold.
The BP   algorithm for proactive Doppler shift compensation is summarized in Algorithm 1.
\begin{algorithm}[b]
\caption{BP for Proactive Doppler Shift Compensation}
\begin{algorithmic}[1]
\STATE  {Initialize:}
Set the initial incoming message parameters $\big[C_{k\arrow j}^{(0)}\big]^{-1}>0$ and
$\eta^{(\tau_{k\rightarrow j}(0))}_{k\arrow j}$ can be arbitrary value
 for all $j\in \mathcal{V}$ and $\{k,j\}\in \E$;
\FOR{  $l \in\{1^{\textrm{st}},2^{\textrm{nd}},\ldots, l^{\textrm{th}}_{\textrm{max}}\}$ iteration }
    \STATE  Vehicle $j$ with $j=1,\cdots,N$ \textbf{in parallel}
    \STATE Compute the outgoing messages $\big[C_{j\arrow i}^{(l)}\big]^{-1}$ and $\eta^{(l)}_{j\arrow i}$ to all neighbors $i\in \B{(j)}$ individually,
    via (\ref{f2fC}) and (\ref{f2f2});
    \STATE Transmit $\big[C_{j\arrow i}^{(l)}\big]^{-1}$ and $\eta^{(l)}_{j\arrow i}$ to each neighbor $i\in \B{(j)}$, separately;
    \STATE With the available  $\big[  C_{j\arrow i}^{(\tau_{j\rightarrow i}(l))}\big]^{-1}$  and $\eta^{(\tau_{k\rightarrow j}(l))}_{j\arrow i}$,  $i$ computes the estimate $\hat {  f}_i^{(l)}$ via
    (\ref{beliefu});
    \STATE \textbf{end parallel}
\STATE    If $\Delta_i < th$,  return current estimate $\hat {f}_i^{(l)}$;
\ENDFOR
\STATE If $\Delta_i > th$, BP does not converge.
\end{algorithmic}
\end{algorithm}

It is well known that the convergence of BP is not guaranteed for topology with loops.
Consequently, the BP  algorithm may either converge or diverge, resulting in unreliable estimates.
Moreover, it is apparent that the  outgoing messages, i.e., (\ref{f2fC}) and (\ref{f2f2}), to different neighbors are different, and thus, huge amount of information is broadcasted in the network. Such problem is especially serious in dense traffic and leads to information network traffic congestion \cite{BroadcastinginVanet,  LowCongestionControl}.

To address the above
problems,  in the next section, we design a novel
distributed  algorithm, which not only guarantees the iterative updating convergence
but also has the property that
the  amount of information exchange among vehicles
 is linear to the traffic density.

\subsection{Design of Linear Scaling BP}
\begin{figure}[t]
  \centering
{\epsfig{file=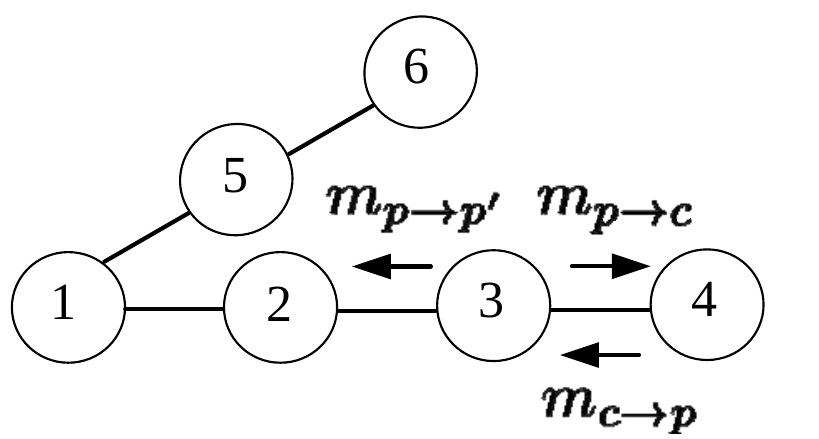,width=2.0in}}
\caption{An example of  probabilistic  graphical model with tree topology.}
\label{chain}
\end{figure}

To get some insights on low communication overhead message passing algorithm, we start by investigating BP in the simplest possible graph:
a tree graphical model.
In this model, BP computes the maximum  likelihood estimate in an
efficient way with convergence guaranteed.
In a tree graphical model as shown in Fig \ref{chain}, for each pair of variables connected by an edge, the variable near the root is named as \emph{parent}, while the other variable is named as \emph{child}.
Then, the messages of BP can be categorized into  two kinds:  one is from parent to child denoted as $m^{(l)}_{p\arrow c}(f_c)$, and the other is from child to parent denoted as $m^{(l)}_{c\arrow p}(f_p)$.
Then, we have the following property.
\begin{myproperty}\label{bbpproperty}
For a tree topology vehicular network with root being the reference vehicle, the BP updating equation (\ref{BPf2vs1}) equals
${m}^{(l)}_{p \arrow c}(f_c)
=  \mathop{\max}_{f_p}
\N(f_{p,c};f_p+f_c,\sigma_{p,c}^2)
 b^{(l-1)}_{p}(f_p)
 $ for message from parent to child,
and
${m}^{(l)}_{c \arrow p}(f_c) $ is a constant
for message from child to parent.
\end{myproperty}
\begin{myproof}
See Appendix \ref{A}.
\end{myproof}

With  Property \ref{bbpproperty}, we can exactly compute the maximum likelihood function for the tree network topology.
Then, we apply it to a network containing loops.
By simply generalizing Property \ref{bbpproperty},  the message in a loopy graph is computed by
$ \tilde{m}^{(l)}_{j \arrow i}(f_i)
=  \mathop{\max}_{f_j}
\N(f_{i,j};f_i+f_j,\sigma_{i,j}^2)
 b^{(l-1)}_{j}(f_j)
$,
and the outgoing message is
$ \tilde{b}^{(l)}(f_i)
 =  \prod_{ j\in\B(i)} \tilde{m}^{( l )}_{ j \arrow i}( f_i) $.

\begin{figure}[t]
  \centering
{\epsfig{file=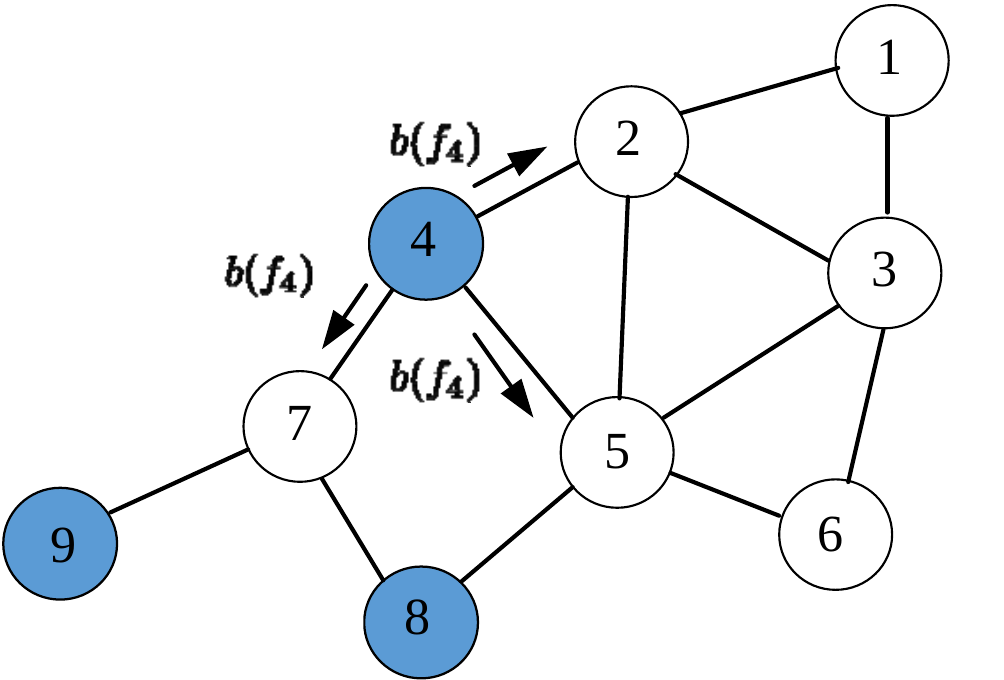,width=2.2in}}
\caption{The same example of a graphical model for a vehicular CPS as shown in Fig. \ref{Graph}.
However,
the  outgoing messages from $4$  to different neighbors are the same in LSBP algorithm.}
\label{BBP}
\end{figure}

For networks with
 random delays and packet drops, the message updating equation can be easily obtained as
\begin{equation}  \label{BBPf2vs1}
\tilde{m}^{(l)}_{j \arrow i}(f_i)
=  \mathop{\max}\limits_{f_j}
\N(f_{i,j};f_i+f_j,\sigma_{i,j}^2)
 b^{(\tau_{j\rightarrow i}(l-1))}_{j}(f_j),
\end{equation}
and the outgoing message is
\begin{equation} \label{BBPbelief}
\tilde{b}^{(l)}(f_i)
 =  \prod_{ j\in\B(i)} \tilde{m}^{(l)}_{ j \arrow i}( f_i).
\end{equation}
Note that (\ref{BBPf2vs1}) differs from the standard BP of (\ref{BPf2vs1-asy}) in that
 each vehicle only transmits  $\tilde{b}^{(l-1)}(f_j)$ to all its neighbors at one time, then $\tilde{m}^{(l)}_{j \arrow i}(f_i)$ is computed at node $i$ and then the belief $\tilde{b}^{(l)}(f_i)$ can be obtained according to (\ref{BBPf2vs1}).
Because  the message need to be transmitted at each iteration equals the number of vehicles,
the proposed method is named as linear scaling BP (LSBP).
Fig. \ref{BBP} denotes the message passing with LSBP.
{Notice that, the message computation equations ((15) and (16))  for each node only depend on message transmitted from neighbor nodes and are independent of network topology, which further implies that there is no need to   construct  a tree topology for message scheduling.}
Next, the explicit message expression of LSBP is computed.

\subsection{Message Computation for Linear Scaling BP}
To start the recursion,
in the first round of message exchange, the initial incoming message is settled as  $b^{(\tau_{j\rightarrow i}(l-1))}_{j}(f_j)=\mathcal{N}( f_j; \mu^{(0)}_{j},  P_{j}^{(0)})$,
with $ P_{j}^{(0)} >0$ and $\mu^{(0)}_{j}$ can be arbitrary value.
Since  $\N(f_{i,j};f_i+f_j,\sigma_{i,j}^2)$ is a Gaussian pdf,
according to (\ref{BBPf2vs1}),  $\tilde{m}^{(1)}_{j \arrow i}(f_i)$ is still a Gaussian function.
In addition, $\tilde{b}^{(1)}(f_i)$, being the product of Gaussian functions in (\ref{BBPbelief}), is also a Gaussian function \cite{du1, chenchen}.
Consequently, in LSBP, during each round of message exchange, all the messages are Gaussian functions, and
only the mean  and the variance need to be exchanged between neighbors.

At this point, we can compute the messages of LSBP at any iteration.
In general, in the $l^{\textrm{th}}$ ($l=2,3,\cdots$) round of message exchange, vehicle $i$ with the available  message
$b^{(\tau_{j\arrow i}(l-1)) }_{ j } (f_j)\propto
\N(f_j;\mu^{(\tau_{j\arrow i}(l-1))}_{j}, P_j^{(\tau_{j\arrow i}(l-1))})$ from its neighbors, computes the outgoing messages via (\ref{BBPf2vs1}).
By putting the explicit expression of
$b^{\tau_{j\arrow i}(l-1)}_{ j } (f_j)$  into
(\ref{BBPf2vs1}) and after some tedious but straightforward  computations, we have
$\tilde{m}^{(l)}_{j \arrow i}(f_i)\propto
\N(f_i; \eta^{(l)}_{j\arrow i}, C_{j\arrow i}^{(l)})$
in which
\begin{equation} \label{f2fCBBP}
  C_{j\arrow i}^{(l)}
 =
  \sigma_{i,j}^2 +
P_j^{(\tau_{j\arrow i}(l-1))},
\end{equation}
and
\begin{eqnarray}\label{f2fvBBP}
\eta^{(l)}_{j\arrow i}
&=&
f_{i,j}+ \mu^{(\tau_{j\arrow i}(l-1))}_{j}.
\end{eqnarray}

Furthermore, during each round of message exchange, each vehicle  computes the belief for $f_i$
via (\ref{BBPbelief}), which can be easily shown to be
$\tilde{b}_{i}^{(l)}(  f_i)\propto\mathcal{N}( f_i;  \mu_i^{(l)},   P_{i}^{(l)})$, with variance
\begin{equation} \label{beliefPBBP}
 P_{i}^{(l)} =   \big[ \sum_{j\in\B(i)}\big[C_{j\arrow i}^{(l)}\big]^{-1}\big]^{-1} ,
\end{equation}
and mean
\begin{equation}\label{beliefuBBP}
 \mu_i^{(l)} =  P_{i}^{(l)} \big\{
 \sum_{j\in\B(i)}
 \big[ C_{j\arrow i}^{(l)}\big]^{-1} \eta^{(l)}_{j\arrow i}\big\}.
\end{equation}
\begin{algorithm}[tp]
\caption{LSBP for Proactive Doppler Shift Compensation}
\begin{algorithmic}[1]
\STATE  {Initialize:}
Set the initial incoming message parameters $P_{j}^{(0)}>0$ and
$\mu_j^{(0)} $ can be arbitrary value for all   $j \in \mathcal{V}$ and $\{j,i\}\in \mathcal{E}$;
\FOR{ $l \in\{1^{\textrm{st}},2^{\textrm{nd}},\ldots, l^{\textrm{th}}_{\textrm{max}}\}$ iteration }
    \STATE  Vehicle $j$ with $j=1,\cdots,N$ \textbf{in parallel}
    \STATE Compute  $\big[  C_{j\arrow i}^{(l)}\big]^{-1}$   and  $\eta^{(l)}_{j\arrow i}$
    via  (\ref{f2fCBBP}) and (\ref{f2fvBBP}) locally at $i$;
    \STATE Compute  $P_i^{(l)}$ and $\mu_i^{(l)}$ with $\big[  C_{j\arrow i}^{(l)}\big]^{-1}$   and  $\eta^{(l)}_{j\arrow i}$ according to (\ref{beliefPBBP}) and (\ref{beliefuBBP}), and $\hat{f}_i^{(l)} = \mu_i^{(l)}$;
    \STATE    If $\Delta_i\triangleq\|\hat{f}_i^{(l)}-\hat{f}_i^{(l-1)}\|<\textrm{th}$,  return current estimate $\hat {  v}_i^{(l)}$;
   \STATE Transmite  $P_i^{(l)}$ and $\mu_i^{(l)}$ to all its neighbors $j\in \B(i)$;
    \STATE \textbf{end parallel}
\ENDFOR
\end{algorithmic}
\end{algorithm}

The updating is iterated between (\ref{f2fCBBP}), (\ref{f2fvBBP}) and
(\ref{beliefPBBP}), (\ref{beliefuBBP}) at each vehicle in parallel.
One way to terminate the iterative algorithm is that all vehicles stop updating when a predefined maximum number of iterations $l_{\textrm{max}}$ is reached.
Since LSBP is convergence guaranteed as proved in the next section, the termination can also be implemented once the algorithm converged.
The LSBP algorithm is summarized in Algorithm 2.


\begin{table}[b]
\center
\caption{Complexity Per Estimation Update}
\label{Tab1}
\begin{tabular}{cl|l}
\hline
\rowcolor[HTML]{EFEFEF}
&\quad \quad \quad \quad  BP   & \quad \quad  LSBP   \\
\hline
&$\big[C_{i\arrow j}^{(\tau_{i\rightarrow j}(l))}\big]^{-1}$, $\eta^{(\tau_{i\rightarrow j}(l))}_{i\arrow j}$, $\forall j\in \B(i)$      &
\quad
$P_{i}^{(l)}$,
       $\mu_{i}^{(l)}$ \\
\hline
\end{tabular}
\label{NoMess}
\end{table}
Table \ref{NoMess} shows messages need to be computed and transmitted at each vehicle at each iteration for BP  and LSBP.
It can be easily concluded that in contrast to BP algorithm, with which the amount of messages need to be computed and transmitted by each vehicle at each iteration is proportional to the number of neighbors, with LSBP each vehicle only needs to compute and transmit one pair of mean and variance to all its neighbors.
Therefore, LSBP is scalable with traffic density.
Moreover, in a limit case where $\mathcal{G}$ is a fully connected graph, i.e., $|\mathcal{B}(i)|=N-1$,
as can be seen from Table \ref{NoMess}, the number of messages exchanged in the network with BP  is $(N-1)N$.
Thus the total number
of messages,  grows quadratically when the
vehicle number $N$ increases, leading to information network
congestion.
While with LSBP it is only $N$.
Therefore, the number of messages involved in BP increases much faster than that with LSBP which leaves the network vulnerable to information congestion.
To get further insights of the  proposed LSBP algorithm, its convergence property is studied in the following section.

\section{Convergence Analysis for LSBP}\label{analysis}
As BP may diverge if the network topology contains circles \cite{du2,du3}, which is often the case in vehicular CPS, BP is not reliable.
In this section, we analytically proved that the proposed LSBP algorithm is convergence guaranteed
with feasible initial values,  and $\mu^{(l)}_{j}$ and $ P_{j}^{(l)} $ converge to the same fixed point respectively even with different initial value pairs  $\mu^{(0)}_{j}$ and $ P_{j}^{(0)}$.
Due to the estimate by LSBP  shown in (\ref{beliefuBBP}) depends on  $P_{i}^{(l)} $ and
$\eta^{(l)}_{j\arrow i}$, we first prove
the convergence of $ P_{i}^{(l)}$ and then $\eta^{(l)}_{j\arrow i}$.

\subsection {Convergence of Message Variance} By substituting (\ref{f2fCBBP}) into (\ref{beliefPBBP}), the updating equation of $P_{i}^{(l)}$ is given by
\begin{equation} \label{beliefP2}
 \big[ P_{i}^{(l)}\big]^{-1} =   \sum_{j\in\B(i)}\big[  \sigma_{i,j}^2 +
P_j^{(\tau_{j\arrow i}(l-1))} \big]^{-1}.
\end{equation}
Let $\bm p^{(l)}$ be  a vector containing of all the message variance at the $l^{\textrm{th}}$ iteration, i.e., $\bm p^{(l)}\triangleq[[P_2^{(l)}]^{-1},[P_3^{(l)}]^{-1},\ldots, [P_N^{(l)}]^{-1}]^T$ and  define an evolution function $ \mathbb{F} $ as $\bm p^{(l+1)}=  \mathbb{F}(\bm p^{(l)}) $.
We will say that a $\bm p^{(0)}>0$ is
a feasible initial value if $\bm p^{(0)}>0$ satisfies
$\mathbb{F}(\bm p^{(0)})\geq \bm p^{(0)})$
or $\mathbb{F}(\bm p^{(0)})\leq \bm p^{(0)}$.
Notice that one easy obtained feasible
$\bm p^{(0)}$ is by setting $\big[ P_{i}^{(l)}\big]^{-1}=0$.
Next, it is shown that the function $\mathbb{F}(\cdot)$ has the following properties
for arbitrary $\bm p^{(0)}>0$.
\begin{myproperty} \label{P-Cov}
The following claims hold with $l\in \{0,1,\cdots\} $:\\
P\ref{P-Cov}-1. Positive limited range: $\mathbb{F}(\bm 0 ) >\mathbb{F}(\bm p^{(l)} )>0$.\\
P\ref{P-Cov}-2. Scalability: $\forall \alpha>1, \alpha \mathbb{F}(\bm p^{(l)})  > \mathbb{F}(\alpha \bm p^{(l)} )$.\\
P\ref{P-Cov}-3. Monotonicity: if $\bm p^{(l)}  \geq \tilde{\bm p}^{(l)}$ then
$\mathbb{F}(\bm p^{(l)} ) \geq \mathbb{F}(\tilde{\bm p}^{(l)})$.
\end{myproperty}
\begin{myproof}
See Appendix \ref{B}
\end{myproof}
Then we can prove the convergence property of the belief variance in LSBP.
\begin{mytheorem}\label{ConConvTh}
With arbitrary feasible initial value $P_i^{(0)}$, the belief variance $ P_i^{(l)}$ of LSBP shown in (\ref{beliefP2})  converges to a unique
fixed point for a specific network topology.
\end{mytheorem}
\begin{myproof}
For arbitrary feasible initial variance $P_i^{(0)} $ after the first round updating, we have 
$\bm p^{(1)}\geq\bm p^{(0)}$ or $\bm p^{(1)}\leq\bm p^{(0)}$.
We first investigate the case when $\bm p^{(1)}\geq\bm p^{(0)}$.
According to P\ref{P-Cov}-3, we have
$\mathbb{F}(\bm p^{(1)})>\mathbb{F}(\bm p^{(0)})$ or equivalently $ \bm p^{(2)} > \bm p^{(1)} $.
Then, the monotonic increasing property of $\bm p^{(l)}$ can be proved by induction following P\ref{P-Cov}-3.
According to P\ref{P-Cov}-1,  $\bm p^{(l)}$ is upper bounded by $\mathbb{F}(\bm 0 )$.
From the monotone convergence theorem \cite{Bertsekas89}, therefore, $\bm p^{(l)}$ is convergence guaranteed.
With the same argument, we can prove that if
$\bm p^{(1)}\leq\bm p^{(0)}$, $\bm p^{(l)}$ is a monotone decreasing positive sequence, which is convergence guaranteed.

In the subsequent, the unique property of the converged $\bm p^{(l)}$ for a    specify network topology is proved by contradiction.
Suppose $\bm p^{\ast}$ and $\tilde{\bm p}^{\ast} $ are two distinctive fixed point, and without loss of generality assume  $\bm p^{\ast} > \tilde{\bm p}^{\ast} $.
Due to the elements in $\bm p$ and $\tilde{\bm p} $ are all positive, there exists $\alpha>1 $ such that
$  \alpha \tilde{\bm p}^{\ast}\geq\bm p^{\ast}  $ and for some particular index $i$ that
\begin{equation}\label{contra1}
\alpha \tilde{P}^{\ast}_i= P^{\ast}_i.
\end{equation}
On the other side, following the definition of fixed point,
we have $\bm p^{\ast}(i)= \mathbb{F}(\bm p^{\ast})(i) \leq \mathbb{F}(\alpha \tilde{\bm p}^{\ast})(i)$, where the inequality comes from the monotonic property (P\ref{P-Cov}-3).
Then following the scalability property (P\ref{P-Cov}-2),
we have
\begin{equation}\label{contra2}
P^{\ast}_j<\alpha \tilde{P}^{\ast}_j.
\end{equation}
Hence, (\ref{contra1}) and (\ref{contra2}) is a contradiction, and
$\bm p^{\ast}$ and $\tilde{\bm p}^{\ast} $ are the same fixed point a specify network topology.
Therefore the element $P_i^{(l)}$ in $\bm p^{(l)}$ converge to a fixed positive value.
This completes the proof.
\end{myproof}

Next, we focus on the convergence property of the estimate $\mu_i^{(l)}$ with the conclusion that $P_i^{(l)}$ has converged.
\subsection {Convergence of Message Mean}
Suppose the converged value of
$P_j^{(l)}$ is $P_j^{\ast}$, then following (\ref{f2fCBBP}), we have
$C_{j\arrow i}^{(l)}=\sigma_{i,j}^2 +
P_j^{\ast}$.
Thus, $C_{j\arrow i}^{(l)}$ is also convergence guaranteed, and then the converged value is denoted by
$C_{j\arrow i}^{\ast}$.
Putting $C_{j\arrow i}^{\ast}$ into (\ref{beliefPBBP}) and substituting the result into (\ref{beliefuBBP}), we have
\begin{equation}\label{dsbp}
\begin{split}
\mu_i^{(l)}
= [\sum_{j\in\B(i)}\big[C_{j\arrow i}^{\ast}\big]^{-1}]^{-1}
\Big\{
 \sum_{j\in\B(i)}\big[
C_{j\arrow i}^{\ast}\big]^{-1}(r_{ij}-\mu_j^{(l-1)})\Big\}.
\end{split}
\end{equation}
In the subsequent, we prove the following theorem for the convergence property of $\mu_i^{(l)}$.
\begin{mytheorem}
For asynchronous updating, with feasible  initial $P_j^{(0)}$,
the mean of LSBP algorithm, i.e., $\mu_i^{(l)}$ in (\ref{dsbp}), converges to a
fixed point irrespective of the network topology.
\end{mytheorem}
\begin{myproof}
Let
$K_{ji} \triangleq  { [\sum_{j\in\B(i)}\big[C_{j\arrow i}^{\ast}\big]^{-1}]^{-1}} {\big[
C_{j\arrow i}^{\ast}\big]^{-1} },
$
and
$\xi_i \triangleq[\sum_{j\in\B(i)}\big[ C_{j\arrow i}^{\ast}\big]^{-1}]^{-1}
\Big\{
 \sum_{j\in\B(i)}\big[
C_{j\arrow i}^{\ast}\big]^{-1}r_{ij}\Big\}
$,
then (\ref{dsbp}) can be expressed as
\begin{equation}\label{beliefu2}
\mu_i^{(l)} =\xi_i -
 \sum_{j\in\B(i)}
 K_{j,i} \mu_j^{(l-1)}.
\end{equation}
Due to the fact that $f_1$ is the reference for pre-compensated frequency shift estimation, thus $\mu_1^{(l)}$ is a constant which is denoted by $\mu_1$, and then  only the convergence of $\mu_2^{(l)},\mu_3^{(l)},\ldots,\mu_N^{(l)}$ needs to be investigated.
Hence, we separate $\mu_1$ from $\sum_{j\in\B(i)},
 K_{j,i} \mu_j^{(l-1)}$ in (\ref{beliefu2}), and the result can be expressed as
\begin{equation}\label{beliefu3}
\mu_i^{(l)} =(\xi_i - K_{1,i}\mu_1\mathbbm{1}_{1,i})
 -
 \sum_{j\in\{\B(i)\setminus 1\}}
\mathbbm{1}_{j,i} K_{j,i} \mu_j^{(l-1)},
\end{equation}
where
$\mathbbm{1}_{j,i}$ is an indicator random variable with $\mathbbm{1}_{j,i}=1$ if $\{j,i\}\in \mathcal{E}$ otherwise is $\mathbbm{1}_{j,i}=0$.

Next, the convergence of $\mu_2^{(l)},\mu_3^{(l)},\ldots,\mu_N^{(l)}$ will be investigated all together.
Define $\bm \mu^{(l)} = [\mu_2^{(l)},\mu_3^{(l)},\ldots,\mu_N^{(l)}]^T$,
and $\bm k_i = [\mathbbm{1}_{2,i} K_{2,i},\mathbbm{1}_{2,i} K_{2,i},\ldots, \mathbbm{1}_{N,i} K_{N,i}]^T$, and then
(\ref{beliefu3}) can be reformulated as
\begin{equation}\label{beliefu4}
\mu_i^{(l)} =(\xi_i - K_{1,i}\mu_1\mathbbm{1}_{1,i})
 -
 \bm k_{i}^T \bm\mu^{(l-1)}.
\end{equation}

Piling up (\ref{beliefu4}) for all $\mu_i$ with the increasing order on $i$, we obtain the updating equation for all $ \bm\mu$ as
\begin{equation}\label{beliefu3}
\bm \mu^{(l)} =\bm \eta
 -
 \bm K\bm\mu^{(l-1)} ,
\end{equation}
where $\bm \eta= [\xi_2 - K_{1,2}\mu_1\mathbbm{1}_{1,2},\xi_2 - K_{1,2}\mu_1\mathbbm{1}_{1,2},\ldots]^T$ and $\bm K$ is an $(N-1)\times (N-1)$ matrix with the $i^{\textrm{th}} $ row of $\bm K$ being $\bm k_i^T$.
According to the definition of $\bm k_i$ above (\ref{beliefu4}), the summation of $\bm k_i$ can be written as
$\sum_{j\in \B(i)\setminus 1}K_{j,i}={\sum_{j\in \B(i)\setminus 1}\big[
 \bm C_{j\arrow i}^{\ast}\big]^{-1} }/{ \sum_{j\in\B(i)}\big[\bm C_{j\arrow i}^{\ast}\big]^{-1}}$.
It is obvious that,
if $1\in \B(i)$,
 $\sum_{j\in \B(i)\setminus 1}K_{j,i}< 1$, and
if $1\not\in \B(i)$,
 $\sum_{j\in \B(i)\setminus 1}K_{j,i}\leq 1$.
Therefore, $\bm K$ is a non-negative matrix having row sums less than or equal to $1$ with at least one row sum less than $1$.
Hence, $\bm K$ is a substochastic matrix.
Consequently, $K$   in  (\ref{beliefu3}) is a non-negative and irreducible substochastic matrix, therefore, $\rho(|\bm K|)=\rho(\bm K)<1$, where
 $\rho(\cdot)$ denotes the spectrum radius of a matrix.
Then (\ref{beliefu3}) is convergence guaranteed \cite{MatrixAnalysis}.
Hence, the convergence of $\mu_i^{(l)}$ in (\ref{dsbp}) is guaranteed irrespective the network topology.
\end{myproof}

\section{Experiment Evaluations}\label{simu}
\begin{figure}[t]
  \centering
{\epsfig{figure=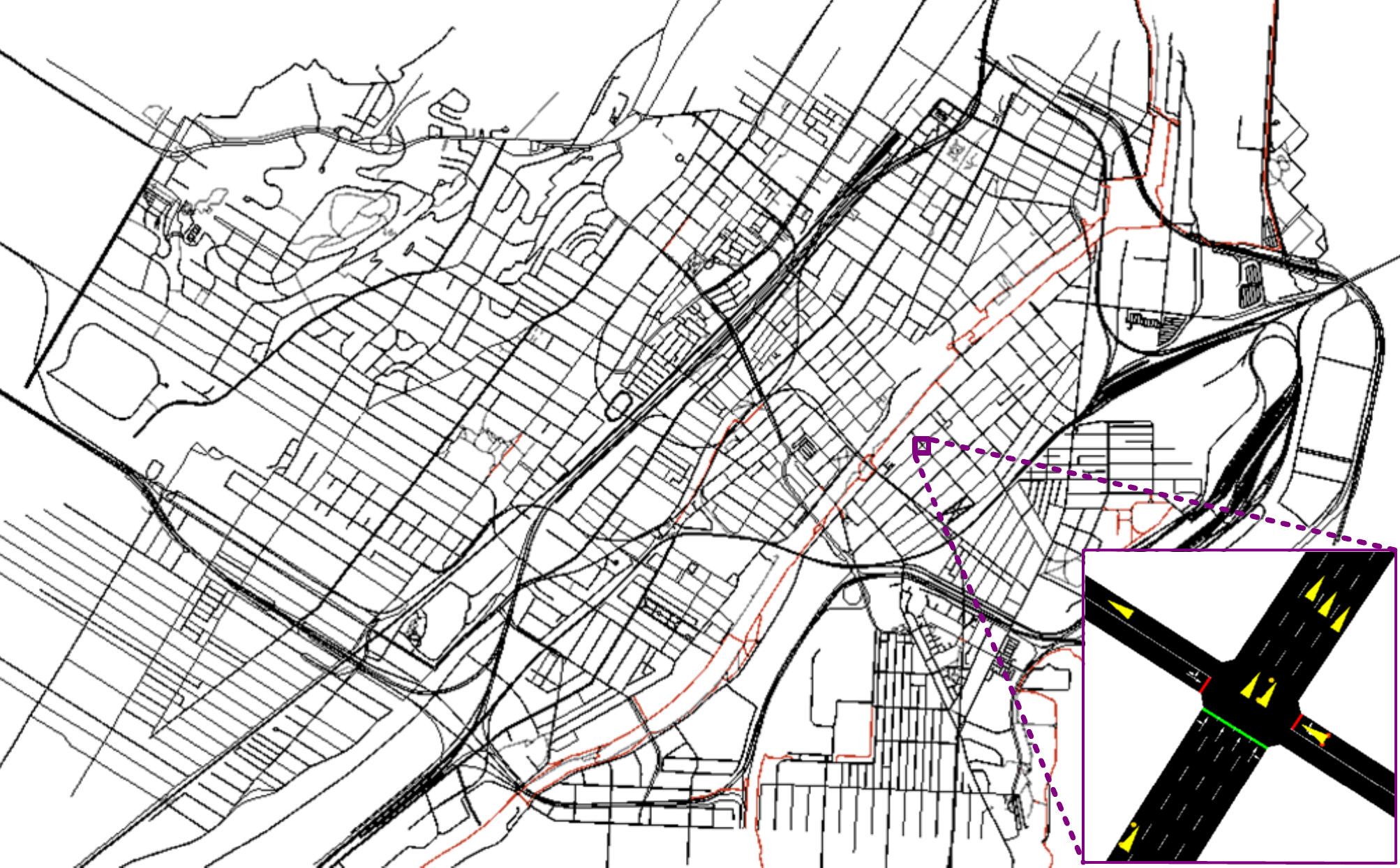,width=2.6in}}
\caption{The street map covering a $3\textrm{km}\times 4\textrm{km}$ area of Montreal from OpenStreetMap \cite{OpenStreetMap}, with $100$ vehicles  generated by SUMO \cite{Sumo}.
The yellow triangles stand for vehicles running on streets.
}
\label{Montreal}
\end{figure}

In this section, realistic data traces and simulation tools are employed to evaluate the proposed algorithm for proactive Doppler shift compensation.
As shown in Fig. \ref{Montreal}, a real street map covering a $3\textrm{km}\times 4\textrm{km}$ area of Montreal is generated from OpenStreetMap \cite{OpenStreetMap}.
Hereafter, and unless stated otherwise, $100$ vehicles are generated on the map by
simulation tool SUMO \cite{Sumo}.
The traffic data generated by SUMO includes vehicular positions, destinations, travelling paths and speeds.
These parameters are also within practical limitations as in Fig. \ref{Montreal}.
For example, vehicle speeds are within the speed limitation of corresponding street.
According to \cite{Kenney:2011Procee}, the communication range of each vehicle is set to be $800$m.
The true Doppler shift between each pair of vehicles within communication range can be computed according to  $f_{i}+f_{j}=v_{i,j}f_0/c$, where $v_{i,j}$ is the relative velocity between vehicles $i$ and $j$, $f_0$ is the carrier frequency, and $c$ is the speed of waves.

In practice, message exchange between vehicles may fail due to
various factors, like separation distance, signal propagation
environment,
received signal strength, transmission power and modulation
rate \cite{Bai:Mobi10}.
In the following experiments, different packet delivery ratio (PDR), which is the ratio of the number of packets successfully delivered to destination compared to the number of packets that have been sent out by the transmitter, is set to show the impacts of packet drop on proposed algorithms.

\begin{figure}[t]
  \centering
{\epsfig{file=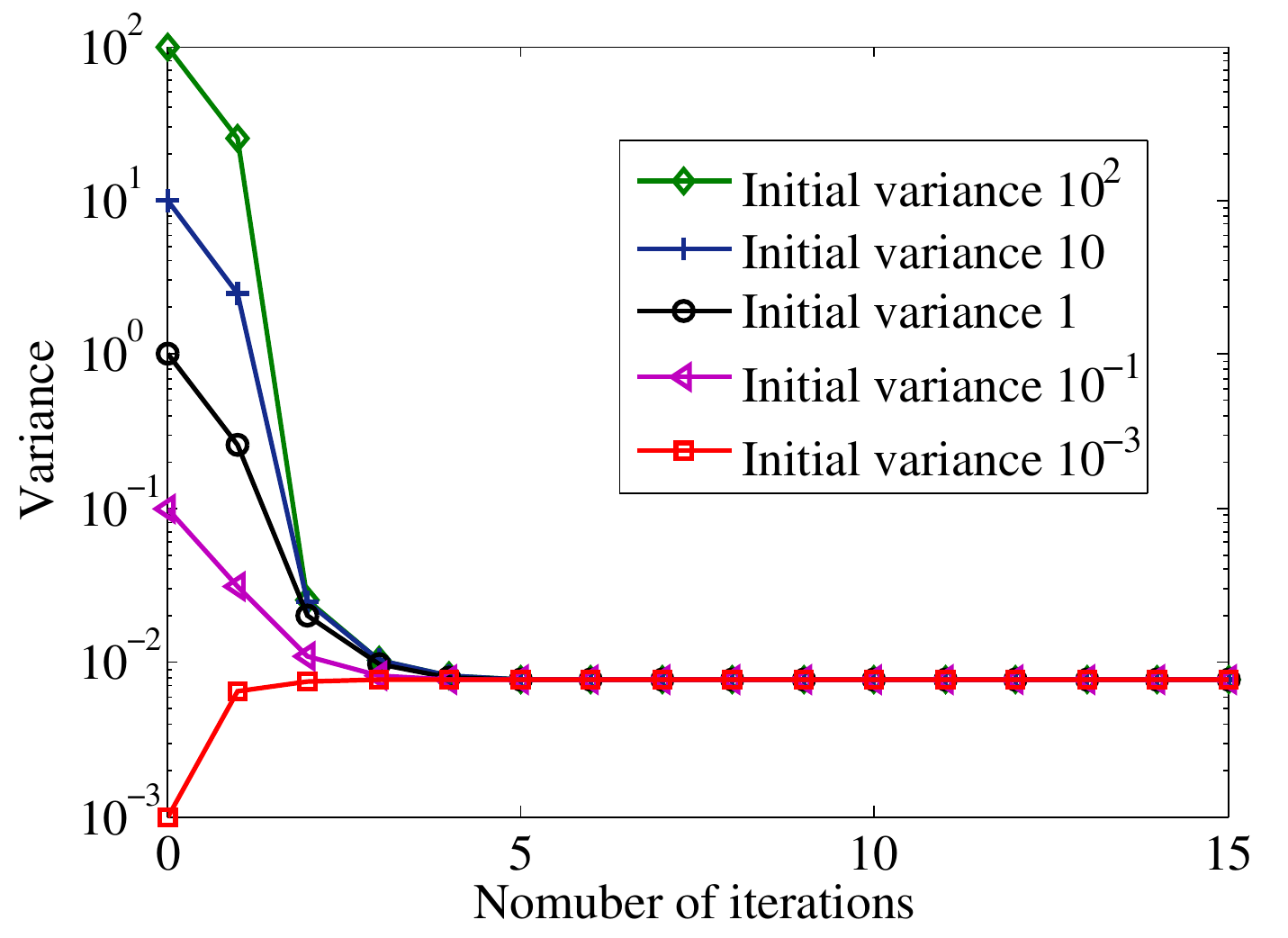,width=2.7in}}
\caption{Convergence property of $\bm P_6^{(l)}$  for different initial values.  }
\label{1234}
\end{figure}

First, the convergence property of $ P_{i}^{(l)}$ as shown in Theorem \ref{ConConvTh} is verified by simulations.
The network topology is randomly generated by SUMO, and PDR is set to be $80\%$.
The initial message variance for each
$ P_{i}^{(0)}$  is set to be $100$, $10$, $1$, $0.1$ and $0.01$, respectively.
The convergence property of $ P_{6}^{(l)}$ is demonstrated in Fig. \ref{1234} as an example.
It is clear that though $ P_{6}^{(l)}$ keeps monotonic increasing or decreasing with different initial values, they converge to the same point.
Thus, the conclusion of Theorem \ref{ConConvTh} is verified by simulations that
with arbitrary feasible initial value, the belief variance $ P_i^{(l)}$ of LSBP shown in (\ref{beliefP2})  converges to a unique fixed point.

\begin{figure}[t]
  \centering
{\epsfig{file=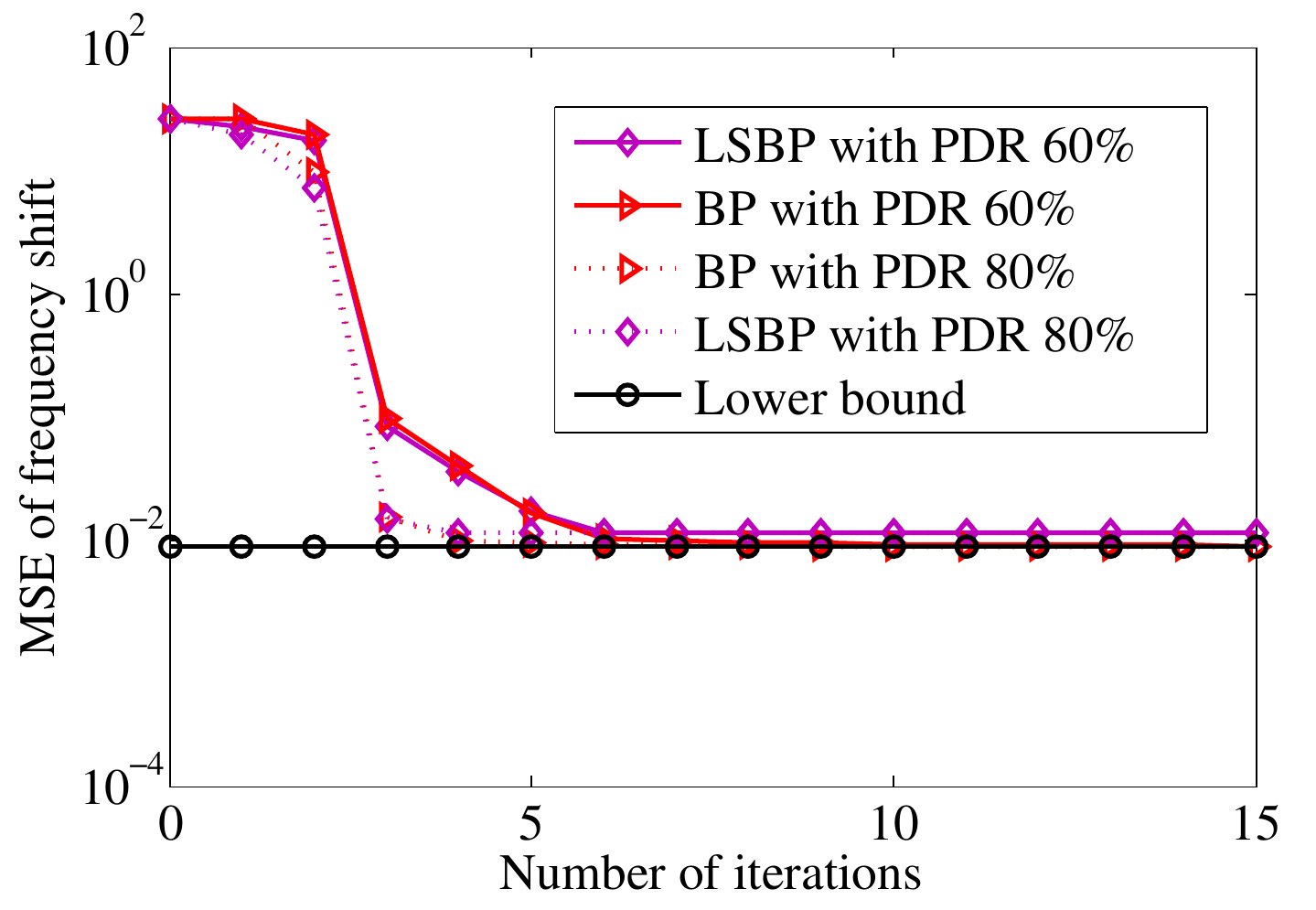,width=2.7in}}
\caption{Accuracy and convergence property of $\hat{f}_i$ under different PDRs.}
\label{MSE-Iter-PDR}
\end{figure}

Next, the accuracy and convergence property of $\hat{f}_i$ is investigated.
Average MSE, defined as
$\frac{1}{N}\sum_{i=1}^{N}\mathbb{E}\{(\frac{\hat{f}_i-f_i}{B})^2\}$,
is adopted as the performance criteria.
Fig. \ref{MSE-Iter-PDR} shows that for different PDRs ($60\%$ and $80\%$), the convergence speeds of BP and LSBP algorithms differ.
Nevertheless, even for PDR as low as $60\%$, both BP and LSBP converge to a fixed estimate point within $10$ iterations, and thus, they are robust to packet drops.
Besides, LSBP has the MSE performance that approaches the CRLB.
Note that BP can also reach the CRLB as shown in Fig. \ref{MSE-Iter-PDR}, but its convergence for loopy topology network is not guaranteed, and its communication overhead is large as shown in Table  \ref{NoMess}.

\begin{figure}[t]
  \centering
{\epsfig{file=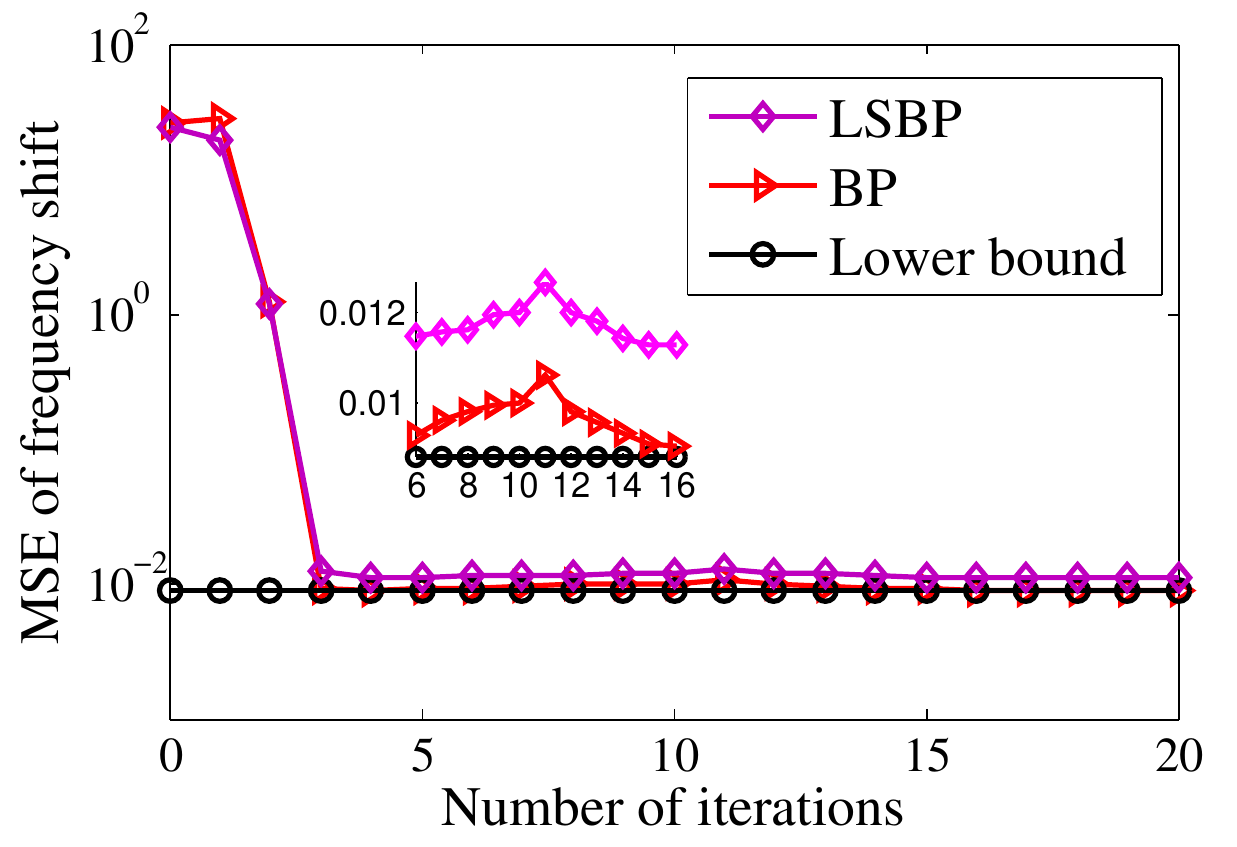,width=2.7in}}
\caption{Adaptive property of proposed algorithms to dynamic  vehicular network.
At iteration $5$, vehicle $4$, $5$, $8$ and $10$  leave the network, and at iterations $10$ and $11$,  new vehicles join the network at former positions of $4$, $5$, $8$ and $10$, respectively.}
\label{MSE-dynamic}
\end{figure}

\begin{figure}[t]
  \centering
{\epsfig{file=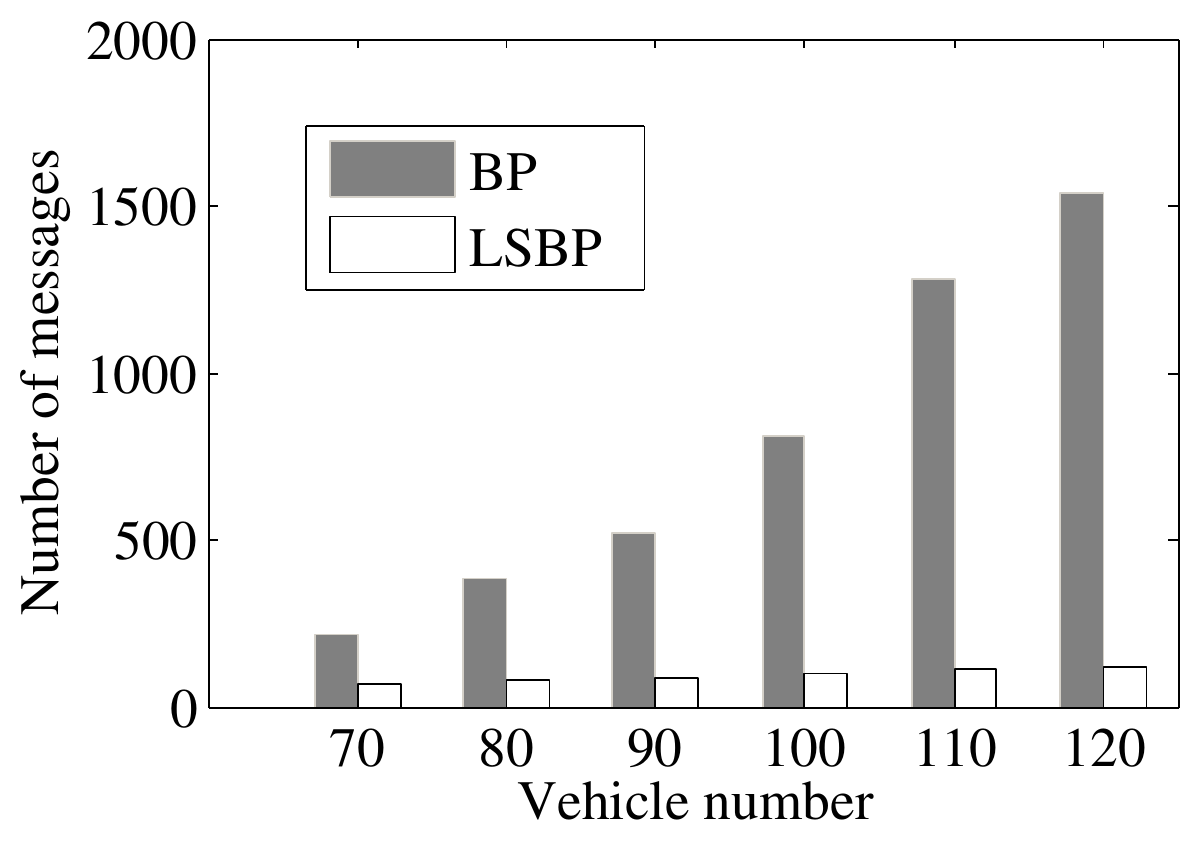,width=2.7in}}
\caption{Comparison on the total number of messages transmitted in a vehicular CPS
at each iteration for BP and LSBP. }
\label{MessageNo}
\end{figure}

\begin{figure}[!t]
  \centering
{\epsfig{file=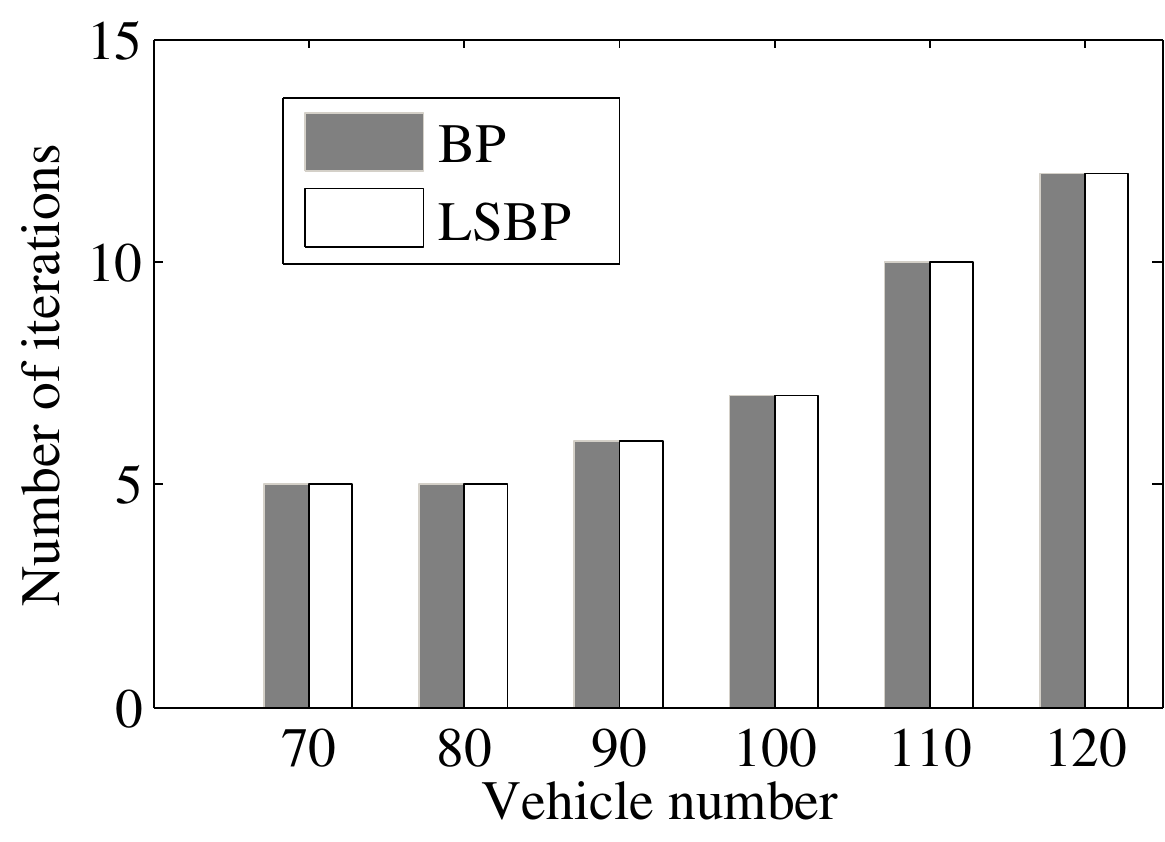,width=2.6in}}
\caption{Iteration numbers upon convergence versus the vehicle number.}
\label{NoIter}
\end{figure}

\begin{figure}[!t]
  \centering
{\epsfig{file=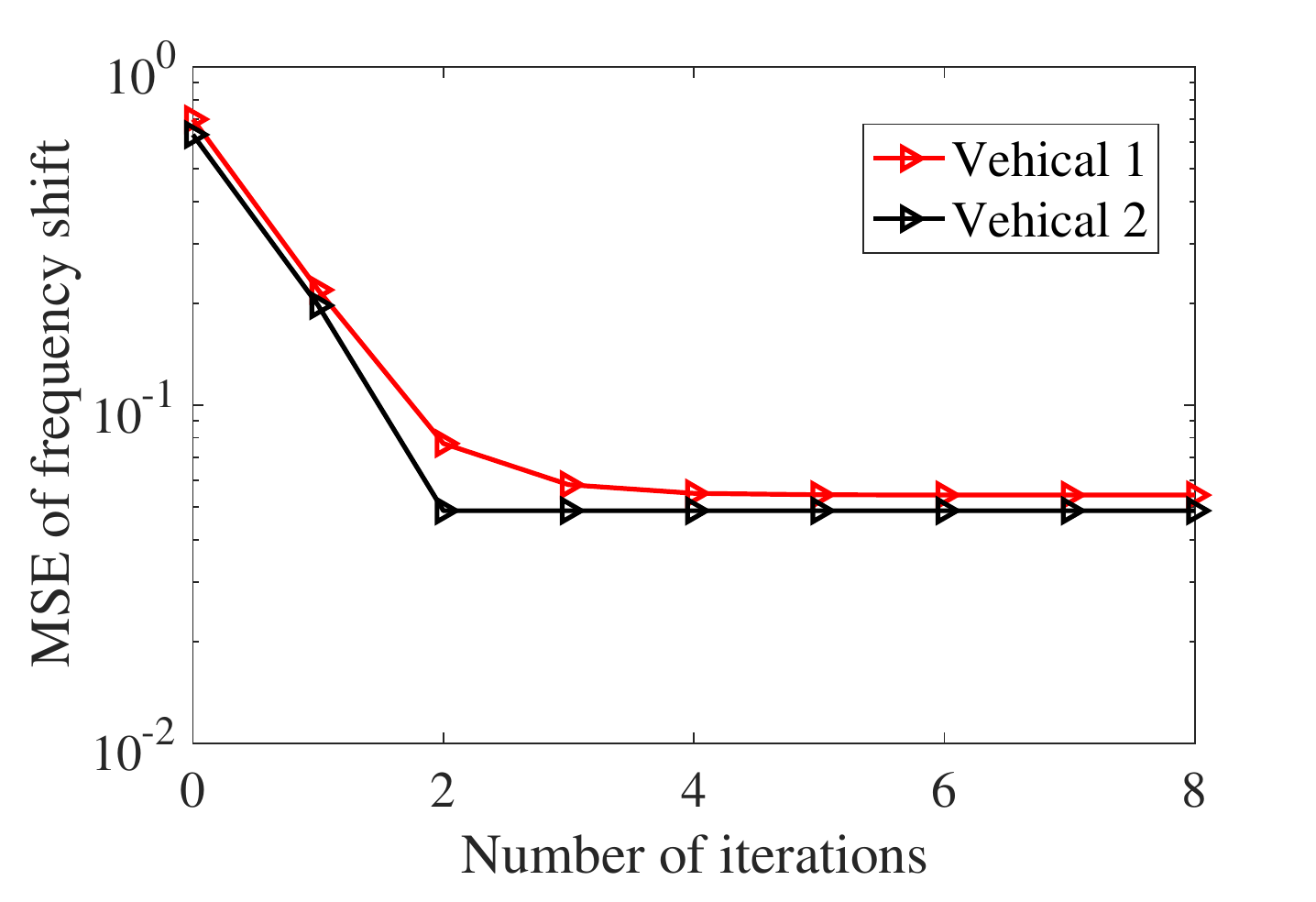,width=3in}}
\caption{Iteration numbers upon convergence versus the vehicle number.}
\label{last}
\end{figure}

Fig. \ref{MSE-dynamic} shows adaptiveness property of the proposed algorithms to the dynamic topology of vehicular networks.
At first, the network topology is the same as that adopted in Fig. \ref{MSE-Iter-PDR}.
At iteration $5$, vehicles $4$, $5$, $8$ and $10$  leave the network, and at iterations $10$ and $11$,  new vehicles join the network at former positions of $4$, $5$, $8$ and $10$, respectively.
It can be seen that the average MSE increases at iteration $6$ due to vehicles' leaving, and it decreases after iteration $11$ because new vehicles join in and bring new measurements.
It is shown that the impact of vehicles' leaving and joining on the performance of BP  and LSBP  is very trivial, and both algorithms are adaptive to topology varying.

In the following, the communication burden imposed by BP and LSBP are analyzed and compared.
First, the total number of messages transmitted among vehicles at each iteration for BP and LSBP is compared in Fig. \ref{MessageNo}.
It is shown that
as the vehicle number increases from $70$ to $120$,
 the number of messages required to be transmitted in BP increases quickly, which may lead to information network congestion \cite{BroadcastinginVanet, Yang2}.
It verifies the analysis of Table \ref{NoMess} that the number of messages increases quadratically for dense network.
In contrast, the number of messages involved in LSBP increases mildly, which, in fact, is simply equal to the number of vehicles.
Next, the number of iterations with BP and LSBP for different scales of vehicles are studied.
As shown in Fig. \ref{NoIter}, for both BP and LSBP algorithms, the iteration number upon convergence increases mildly with the increasing number of vehicles.
This property makes sure that proactive frequency compensation can be achieved within limited time in vehicular CPS.
From both Figs. \ref{MessageNo} and \ref{NoIter}, we can conclude
that LSBP has much lower communication overhead compared with BP and is much preferred in dense traffic networks.
{Furthermore,   the LSBP algorithm proposed does not need a control center and control channel to coordinate the distributed computing or perform scheduling.
Each computation is performed locally and information is only needed to be transmitted to the direct neighbors.
The overhead is very small compared with communication information that not only includes safety information but also  transmission data for social network and entertainment such as video/voice data \cite{1,2,3,4,5,6}.}

{
Next, we show that the overhead of LSBP is reasonable and practical.
We adopt double-precision floating-point format which uses $8$ bytes to represent decimal fraction.
Since each time only two real value scalars (mean and variance) need to be transmitted at each node, information needs to be transmitted are $16$ bytes.
According to the empirical measurement in \cite{Bai:Mobi10}, A $300$-Byte packet  takes $0.4$ms transmission time using $6$ Mbps (QPSK) in vehicular networks.
It is evident that time needed to transmit
 the above $16$ bytes
 in each iteration is smaller than $0.4$ms.
Take a network with $100$ vehicles as an example,
as shown in Fig. 11, the average number of iteration is around $7$.
Since the proposed method is a parallel algorithm,  the transmission time is smaller than $0.4\times 7=2.8$ms, which is a acceptable overhead for wireless communication.}

{At last, we show that  even for  highway environment where the network topology is linear and the nodes are sparsely connected the proposed LSBP algorithm still works.
We assume $10$ vehicles scattered as a line in a highway and each vehicle's speed is generated by SUMO simulations. It is assumed that each vehicle can only communicate with its front and back neighbors.
The convergence performance of estimation MSE is given in Fig. \ref{last}.
It is shown that it is less accurate than the dense network topology case that we considered in Fig. 8.
But the performance is accurate enough for further data detection.}

\section{Conclusions}\label{conclusion}
In this paper, an algorithm for proactive Doppler shift compensation has been proposed to enhance the reliability of safety information sharing in vehicular cyber-physical systems.
Probabilistic graphical model has been
incorporated to reveal the conditional independence property of the pre-compensated frequency offset at each vehicle.
In this distributed message passing algorithm, named as linear scaling belief propagation (LSBP), the communication overhead is linear scaling with the network density.
Analytical analysis has been conducted to rigorously prove that the the proposed algorithm is convergence guaranteed wit feasible initial values even for systems with packet drops and random delays.
Though LSBP only requires local information at each vehicle, simulations based on real map and transportation data have verified that LSBP achieves almost the optimal frequency compensation accuracy with an error approaching the Cram\'{e}r-Rao lower bound.
Simulations also show that the number of exchanged messages linearly scales with the number of vehicles, and the iteration number upon convergence increases mildly, and thus, implementing LSBP imposes tolerable communication overhead.

\appendices
\section{Proof of Property \ref{bbpproperty} }\label{A}
Observe that for a tree topology computations of ${m}^{(l)}_{p \arrow c}(f_c)$
and ${m}^{(l)}_{c \arrow p}(f_p)$ are independent, so we compute them separately.

First, we compute ${m}^{(l)}_{c \arrow p}(f_c)$ by starting from a variable $c$ without any child.
According to (\ref{BPf2vs1}), we have
\begin{eqnarray} \label{compute1}
m^{(l)}_{c \arrow p}(f_{p})
 &= & \mathop{\max}\limits_{f_c}
\psi_{p,c}(f_{p},f_{c})
  \nonumber\\
 &= &  \mathop{\max}\limits_{f_c}
\N(r_{p,c};f_{p}+f_{c},\sigma_{p,c}^2)  \nonumber\\
 &=&
\mathop{\max}\limits_{f_c}
\N(f_{c} ;r_{p,c}-f_{p},\sigma_{p,c}^2)
\nonumber\\
 &=& {\sigma_{p,c}^{-2}}/{\sqrt{2\pi}},
\end{eqnarray}
where the third equation is from  $\N(x;\mu,\sigma^2)=\N(\mu;x,\sigma^2)$, and the fourth equation comes from the fact that maximum value of Gaussian PDF only relates to its variance.
Next, we compute the message from $p$ to its parent $p^{\prime}$.
According to (\ref{BPf2vs1}), we have
\begin{eqnarray} \label{compute2}
m^{(l)}_{p \arrow p^{\prime}}(f_{p^{\prime}})
 &= & \mathop{\max}\limits_{f_p}
\psi_{p^{\prime},p}(f_{p^{\prime}},f_p)
\prod_{c\in\B(p)\setminus p^{\prime}}\!\!\! m^{(l-1)}_{c \arrow p}(f_p)
 \nonumber\\
 &\propto&
\mathop{\max}\limits_{f_p}
\N(f_{p} ;f_{p^{\prime}}+r_{p^{\prime},p},\sigma_{p^{\prime},p}^2)
 \nonumber\\
 &=& {\sigma_{p^{\prime},p}^{-2}}/{\sqrt{2\pi}},
\end{eqnarray}
where the first equation is due to $m^{(l)}_{c \arrow p}(f_{p})$ is a constant.
Then, by induction, we have ${m}^{(l)}_{c \arrow p}(f_c)
$ is constant for all messages from child to parent.
Therefore, this kind of message can be omitted for computation and transmission,
and only $m^{(l)}_{p \arrow c}(f_{(c)})$ needs to  be computed.

Following (\ref{BPf2vs1}), we obtain
\begin{eqnarray} \label{compute1}
m^{(l)}_{p \arrow c}(f_c)
&= &\mathop{\max}\limits_{f_p}
\psi_{p,c}(f_{p},f_{c})\prod_{p^{\prime}\in\B(p)\setminus c}\!\!\! m^{(l-1)}_{p^{\prime} \arrow p}(f_p)
 \nonumber\\
\!\!\!\!&= &\!\!\!\!
 \mathop{\max}\limits_{f_p}
\psi_{p,c}(f_p,f_c)
 b^{(l-1)}_{p}(f_p).
\end{eqnarray}
This completes the proof.

\section{Proof of Property \ref{P-Cov} }\label{B}
We first prove P\ref{P-Cov}-1.
Since $\sigma_{i,j}^2 >0$ and $P_j^{(0)}>0$, according to (\ref{beliefP2}), it is obvious that $[P_i^{(1)}]^{-1}>0$.
Then, it can be easily proved by induction that for arbitrary $l$,  $ \big[ P_{i}^{(l)}\big]^{-1}>0$, and thus $\bm p^{(l+1)} = \mathbb{F}(\bm p^{(l)} )>0$.
Then, if $\bm p^{(0)}>0$, by induction we have $\bm p^{(l)} >0$ for  $l\in \{0, 1, 2,\ldots\}$.
Furthermore, according to (\ref{beliefP2}) it is shown that $ \big[ P_{i}^{(l)}\big]^{-1}$ is a monotonic decreasing function with respect to $P_j^{(\tau_{j\arrow i}(l-1))}$.
As $ \big[ P_{i}^{(l)}\big]^{-1}>0$ or
equivalently
$ P_{i}^{(l)}>0$, we have
$\big[ P_{i}^{(l)}\big]^{-1} <   \sum_{j\in\B(i)}\big[  \sigma_{i,j}^2 +
0 \big]^{-1}$
or equivalently
$\mathbb{F}(\bm 0 ) >\mathbb{F}(\bm p^{(l)} )>0$.
Hence, P\ref{P-Cov}-1 is proved.

Next, we prove P\ref{P-Cov}-2.
Let $ \mathbb{F}_i(\bm p^{(l)}) $ denote the $i^{\textrm{th}} $ element in $\mathbb{F}(\bm p^{(l)})$, then according to (\ref{beliefP2}), for arbitrary $\alpha>1$, we have
\begin{equation} \label{compare1}
\alpha  \mathbb{F}_i(\bm p^{(l)})
= \alpha \sum_{j\in\B(i)}\big[  \sigma_{i,j}^2 +
P_j^{(l)} \big]^{-1}.
\end{equation}
Besides, the corresponding $i^{\textrm{th}}$ element in $\mathbb{F}(\alpha\bm p)$ is given by
\begin{equation} \label{compare2}
 \mathbb{F}_i(\alpha\bm p^{(l)})
= \sum_{j\in\B(i)}\big[  \sigma_{i,j}^2 +
\frac{ P_j^{(l)}}{\alpha} \big]^{-1}.
\end{equation}
Computing (\ref{compare1})-(\ref{compare2}), we have
\begin{equation} \label{compare3}
\begin{split}
&\alpha  \mathbb{F}_i(\bm p^{(l)})
 - \mathbb{F}_i(\alpha\bm p^{(l)}) \\
=&
\alpha \sum_{j\in\B(i)}\big\{\big[  \sigma_{i,j}^2 +
P_j^{(l)} \big]^{-1}
- \big[{\alpha}  \sigma_{i,j}^2 +
 { P_j^{(l)}} \big]^{-1}\big\}.
\end{split}
\end{equation}
As $\alpha>1$ and $\sigma_{i,j}^2>0$,
it can be concluded that in (\ref{compare3}), $\alpha  \mathbb{F}_i(\bm p^{(l)})
 - \mathbb{F}_i(\alpha\bm p^{(l)})>0 $.
The above inequality is satisfied for arbitrary $i$,
so we have $\forall \alpha>1, \alpha \mathbb{F}(\bm p)  > \mathbb{F}(\alpha \bm p )$.
Thus, the scalability is proved.

At last, we prove the monotonic property (P\ref{P-Cov}-3).
Denote $ \bm p^{(l)} =
[[P_2^{(l)}]^{-1},[P_3^{(l)}]^{-1},\ldots, [P_N^{(l)}]^{-1}]$ and
$ \tilde{\bm p}^{(l)}=
[[\tilde{P}_2^{(l)}]^{-1},\\\quad [\tilde{P}_3^{(l)}]^{-1},\ldots, [\tilde{P}_N^{(l)}]^{-1}]$.
If $\bm p^{(l)}  \geq \tilde{\bm p}^{(l)}$, we have
$\sum_{j\in\B(i)}\big[  \sigma_{i,j}^2 +
P_j^{(l)} \big]^{-1}
\geq
\sum_{j\in\B(i)}\big[  \sigma_{i,j}^2 +
\tilde{P}_j^{(l)} \big]^{-1} $.
Then, according to (\ref{beliefP2}),
$\big[ P_{i}^{(l+1)}\big]^{-1} \geq \big[ \tilde{P}_{i}^{(l+1)}\big]^{-1}$.
Therefore, $\mathbb{F}(\bm p^{(l)} ) \geq \mathbb{F}(\tilde{\bm p}^{(l)})$. The monotonic property is proved.


\begin{thebibliography}{10}
\providecommand{\url}[1]{#1}
\csname url@samestyle\endcsname
\providecommand{\newblock}{\relax}
\providecommand{\bibinfo}[2]{#2}
\providecommand{\BIBentrySTDinterwordspacing}{\spaceskip=0pt\relax}
\providecommand{\BIBentryALTinterwordstretchfactor}{4}
\providecommand{\BIBentryALTinterwordspacing}{\spaceskip=\fontdimen2\font plus
\BIBentryALTinterwordstretchfactor\fontdimen3\font minus
  \fontdimen4\font\relax}
\providecommand{\BIBforeignlanguage}[2]{{%
\expandafter\ifx\csname l@#1\endcsname\relax
\typeout{** WARNING: IEEEtran.bst: No hyphenation pattern has been}%
\typeout{** loaded for the language `#1'. Using the pattern for}%
\typeout{** the default language instead.}%
\else
\language=\csname l@#1\endcsname
\fi
#2}}
\providecommand{\BIBdecl}{\relax}
\BIBdecl

\bibitem{Murray-control}
R.~M. Murray, ``Recent research in cooperative control of multivehicle
  systems,'' \emph{J. Dynamic Syst., Meas. Control}, vol. 129, no.~5, pp.
  571--583, 2007.

\bibitem{CPS-Connect}
{U.S. Department of Transportation}, ``Transforming transportation through
  connectivity-its strategic research plan, 2010-2014,''
  \url{http://www.its.dot.gov/strategicplan/}.

\bibitem{Kenney:2011Procee}
J.~Kenney, ``Dedicated short-range communications ({DSRC}) standards in the
  united states,'' \emph{Proceedings of the IEEE}, vol.~99, no.~7, pp.
  1162--1182, July 2011.

\bibitem{wang2016smart}
Q.~Wang, X.~Liu, J.~Du, and F.~Kong, ``Smart charging for electric vehicles: A
survey from the algorithmic perspective,'' \emph{IEEE Communications Surveys
	\& Tutorials}, vol.~18, no.~2, pp. 1500--1517, 2016.

\bibitem{2}
F.~Dressler, H.~Hartenstein, O.~Altintas, and O.~K. Tonguz, ``Inter-vehicle
  communication: Quo vadis,'' \emph{IEEE Communications Magazine}, vol.~52,
  no.~6, pp. 170--177, 2014.

\bibitem{1}
G.~Araniti, C.~Campolo, M.~Condoluci, A.~Iera, and A.~Molinaro, ``{LTE} for
  vehicular networking: a survey,'' \emph{IEEE Communications Magazine},
  vol.~51, no.~5, pp. 148--157, 2013.

\bibitem{3}
A.~Vinel, ``{3GPP} {LTE} versus {IEEE} 802.11 p/{WAVE}: Which technology is
  able to support cooperative vehicular safety applications?'' \emph{IEEE
  Wireless Communications Letters}, vol.~1, no.~2, pp. 125--128, 2012.

\bibitem{4}
A.~Z. A.~Bazzi, B. M.~Masini, ``Performance analysis of {V2V} beaconing using
  {LTE} in direct mode with full duplex radios,'' \emph{IEEE Wireless
  Communications Letters}, vol.~4, no.~6, pp. 685--688, 2015.

\bibitem{5}
A.~Bazzi, A.~Zanella, and B.~M. Masini, ``An {OFDMA} based {MAC} protocol for
  next generation {VANETs},'' \emph{IEEE Transactions on Vehicular Technology},
  vol.~4, no.~6, pp. 790--804, 2007.

\bibitem{6}
K.~Zheng, Q.~Zheng, P.~Chatzimisios, and Y.~Z. W.~Xiang, ``Heterogeneous
  vehicular networking: A survey on architecture, challenges, and solutions,''
  \emph{IEEE Communication Surveys \& Tutorials}, vol.~17, no.~4, pp.
  2377--2396, 2015.

\bibitem{V2V-PIEEE}
C.~Mecklenbrauker, A.~Molisch, J.~Karedal, F.~Tufvesson, A.~Paier, L.~Bernado,
  T.~Zemen, O.~Klemp, and N.~Czink, ``Vehicular channel characterization and
  its implications for wireless system design and performance,''
  \emph{Proceedings of the IEEE}, vol.~99, no.~7, pp. 1189--1212, July 2011.

\bibitem{JWChen}
J.~Chen, Y.-C. Wu, S.~Ma, and T.-S. Ng, ``Joint {CFO} and channel estimation
  for multiuser {MIMO-OFDM} systems with optimal training sequences,''
  \emph{{IEEE} Trans. Signal Process.}, vol.~56, no.~8, pp. 4008--4019, Aug.
  2008.

\bibitem{YiqingZhou}
Y.~Zhou, J.~Wang, and M.~Sawahashi, ``Downlink transmission of broadband
  {OFCDM} systems-part {II}: effect of {Doppler} shift,'' \emph{{IEEE} Trans.
  Commun.}, vol.~54, no.~6, June 2006.

\bibitem{zhou}
Y.~Zhou, ``Radio environment map based maximum a posteriori {D}oppler shift
  estimation for lte-r,'' in \emph{2014 International Workshop on High Mobility
  Wireless Communications}, 2014, pp. 5--5.

\bibitem{cai2010cfo}
K.~Cai, X.~Li, J.~Du, Y.-C. Wu, and F.~Gao, ``{CFO} estimation in ofdm systems
under timing and channel length uncertainties with model averaging,''
\emph{IEEE Transactions on Wireless Communications}, vol.~9, no.~3, pp.
970--974, 2010.



\bibitem{chen2015signal}
S.~Chen, R.~Varma, A.~Singh, and J.~Kova{\v{c}}evi{\'c}, ``Signal recovery on
  graphs: Fundamental limits of sampling strategies,'' \emph{arXiv preprint
  arXiv:1512.05405}, 2015.

\bibitem{chen2015discrete}
S.~Chen, R.~Varma, A.~Sandryhaila, and J.~Kovacevic, ``Discrete signal
  processing on graphs: Sampling theory,'' \emph{Signal Processing, IEEE
  Transactions on}, vol.~63, no.~24, pp. 6510--6523, 2015.

\bibitem{BroadcastinginVanet}
O.~Tonguz, N.~Wisitpongphan, F.~Bai, P.~Mudalige, and V.~Sadekar,
  ``Broadcasting in {VANET},'' in \emph{2007 Mobile Networking for Vehicular
  Environments}, May 2007, pp. 7--12.

\bibitem{LowCongestionControl}
L.~Gan, A.~Walid, and S.~Low, ``Energy-efficient congestion control,'' in
  \emph{Proceedings of the 12th ACM SIGMETRICS/PERFORMANCE Joint International
  Conference on Measurement and Modeling of Computer Systems}, 2012, pp.
  89--100.

\bibitem{du2016convergence}
J.~Du, S.~Ma, Y.-C. Wu, S.~Kar, and J.~M. Moura, ``Convergence analysis of
  distributed inference with vector-valued gaussian belief propagation,''
  \emph{arXiv preprint arXiv:1611.02010}, 2016.

\bibitem{du2013network}
J.~Du and Y.-C. Wu, ``Network-wide distributed carrier frequency offsets
  estimation and compensation via belief propagation,'' \emph{IEEE Transactions
  on Signal Processing}, vol.~61, no.~23, pp. 5868--5877, 2013.

\bibitem{du2013distributed}
------, ``Distributed clock skew and offset estimation in wireless sensor
  networks: asynchronous algorithm and convergence analysis,'' \emph{IEEE
  Transactions on Wireless Communications}, vol.~12, no.~11, pp. 5908--5917,
  2013.

\bibitem{Nature-Bat}
W.~Metzner, ``A possible neuronal basis for {Doppler}-shift compensation in
  echo-locating horseshoe bats.'' \emph{Nature}, vol.~46, no.~2, pp. 529--532,
  1989.

\bibitem{DopplorSpeed}
S.~Barnwal, R.~Barnwal, R.~Hegde, R.~Singh, and B.~Raj, ``Doppler based speed
  estimation of vehicles using passive sensor,'' in \emph{2013 IEEE
  International Conference on Multimedia and Expo Workshops (ICMEW),}, July
  2013, pp. 1--4.

\bibitem{PointEst}
E.~Lehmann and G.~Caselle, \emph{Theory of Point Estimation}.\hskip 1em plus
  0.5em minus 0.4em\relax New York: Springer Texts in Statistics, 1998.

\bibitem{Yang}
Y.~Yang, S.~Kar, and P.~Grover, ``Graph codes for distributed instant message
  collection in an arbitrary noisy broadcast network,'' \emph{IEEE Transactions
  on Information Theory}, 2017.

\bibitem{pmudu}
J.~Du, S.~Ma, Y.-C. Wu, and H.~V. Poor, ``Distributed hybrid power state
  estimation under pmu sampling phase errors,'' \emph{IEEE Transactions on
  Signal Processing}, vol.~62, no.~16, pp. 4052--4063, 2014.

\bibitem{robust}
T.~Muller and H.~Rohling, ``Channel coding for narrow-band rayleigh fading with
  robustness against changes in doppler spread,'' \emph{{IEEE} Trans. Commun.},
  vol.~45, no.~2, pp. 148--151, 1997.

\bibitem{yang2017coding}
Y.~Yang, P.~Grover, and S.~Kar, ``Coding method for parallel iterative linear
  solver,'' \emph{arXiv preprint arXiv:1706.00163}, 2017.

\bibitem{Bai:Mobi10}
F.~Bai, D.~D. Stancil, and H.~Krishnan, ``Toward understanding characteristics
  of dedicated short range communications ({DSRC}) from a perspective of
  vehicular network engineers,'' in \emph{Proceedings of the Sixteenth Annual
  International Conference on Mobile Computing and Networking, MobiCom '10},
  2010, pp. 329--340.

\bibitem{du1}
J.~Du, S.~Kar, and J.~M.~F. Moura, ``Distributed convergence verification for
  {G}aussian belief propagation,'' \emph{accepted by IEEE Global Conference
  on Signal and Information Processing}, 2017.

\bibitem{du2}
J.~Du, S.~Ma, Y.-C. Wu, S.~Kar, and J.~M.~F. Moura, ``Convergence analysis of
  belief propagation for pairwise linear {G}aussian models,'' in \emph{accepted
  by IEEE Asilomar Conference on Signals, Systems, and Computers}, 2017.

\bibitem{du3}
J.~Du, S.~Ma, Y.-C. Wu, S.~Kar, and J.~M.~F. Moura, ``Convergence analysis of
  the information matrix in {G}aussian belief propagation,'' in \emph{IEEE
  International Conference on Acoustics, Speech and Signal Processing}, 2017.

\bibitem{MatrixAnalysis}
R.~A. Horn and C.~R. Johnson, \emph{Matrix Analysis}, 2nd~ed.\hskip 1em plus
  0.5em minus 0.4em\relax Cambridge University Press, 2012.

\bibitem{OpenStreetMap}
{Open Street Map}, \url{http://www.openstreetmap.org/}, [Online; accessed
  20-May-2015].

\bibitem{Sumo}
{DLR-Institute of transportation systems}, ``Sumo-simulation of urban
  mobility,'' \url{http://dlr.de/ts/sumo}, [Online; accessed 20-May-2015].

\bibitem{Yang2}
Y.~Yang, P.~Grover, and S.~Kar, ``Computing linear transformations with
  unreliable components,'' \emph{IEEE Transactions on Information Theory},
  2017.

\bibitem{chenchen}
S.~Chen, A.~Sandryhaila, J. M. F. Moura, and J.~Kovacevic, “Signal recovery on graphs: Variation minimization”, \emph{IEEE Trans. Signal Process}., 2015.
\end{thebibliography}
\end{document}